\documentclass[a4paper,fleqn,usenatbib]{mnras}

\usepackage{mathptmx}

\usepackage[T1]{fontenc}
\usepackage{ae,aecompl}

\usepackage{tablefootnote}
\usepackage{graphicx}	
\usepackage{amsmath}	
\usepackage{amssymb}	

\newcommand{\angstrom}{\mbox{\normalfont\AA}}

\title[Feasibility of debris ring transits]{Feasibility of the debris ring transit method for the solar-like star HD 107146 by an occulted galaxy}

\author[van Sluijs $\&$ Vaendel et al.]{
L. van Sluijs,$^{1}$\thanks{E-mail: vansluijs@strw.leidenuniv.nl (LvS); vaendel@strw.leidenuniv.nl (DAJHV)}
D.A.J.H. Vaendel,$^{1}$\footnotemark[1]
\newauthor{
B.W. Holwerda,$^{1,2}$
M.A. Kenworthy,$^{1}$ and
G. Schneider$^{3}$ 
}
\\
$^{1}$University of Leiden, Sterrenwacht Leiden, Niels Bohrweg 2, NL-2333 CA Leiden, The Netherlands\\
$^{2}$Department of Physics and Astronomy, 102 Natural Science Building, University of Louisville, Louisville KY 40292, USA\\
$^{3}$Steward Observatory, The University of Arizona, 933 North Cherry Avenue, Tucson, AZ 85721, USA
}

\date{Accepted XXX. Received YYY; in original form ZZZ}

\pubyear{2018}

\begin{document}
\label{firstpage}
\pagerange{\pageref{firstpage}--\pageref{lastpage}}
\maketitle

\begin{abstract}
Occulting galaxy pairs have been used to determine the transmission and dust composition within the foreground galaxy. Observations of the nearly face-on ring-like debris disk around the solar-like star HD 107146 by HST/ACS in 2004 and HST/STIS in 2011 reveal that the debris ring is occulting an extended background galaxy over the subsequent decades. Our aim is to use 2004 HST observations of this system to model the galaxy and apply this to the 2011 observation in order to measure the transmission of the galaxy through the outer regions of the debris disk. We model the galaxy with an exponential disk and a S\'{e}rsic pseudo-bulge in the V- and I-band, but irregularities due to small scale structure from star forming regions limits accurate determination of the foreground dust distribution. We show that debris ring transit photometry is feasible for optical depth increases of $\Delta \tau \geq$ 0.04 ($1 \sigma$) on tens of au scales -- the width of the background galaxy -- when the 2011 STIS data are compared directly with new HST/STIS observations, instead of the use of a smoothed model as a reference. 
\end{abstract}

\begin{keywords}
Stars: individual HD 107146 -- Techniques: high angular resolution -- Methods: data analysis -- Circumstellar matter -- occultations -- Galaxy: general
\end{keywords}



\section{Introduction}

Debris disks may contain circumstellar belts of dust, analogous to the Solar System's asteroid and Kuiper belts. These disks exist around both young and more evolved main sequence stars. For the youngest, pre-main sequence, stars, the disk is considered to be a remnant of the protoplanetary disk, containing both gas and dust. In the disks of more evolved stars the gas is mostly depleted. The dust in the disks is thought to be continuously replenished by collisions of larger bodies, such as planetesimals \citep{Wyatt2008}. 

Studying a star's debris disk can provide clues on the evolution of its planetesimal belts. Furthermore, the morphology of the debris disk can be gravitationally influenced by planets \citep{Wyatt2008, Pearce2015}. For this reason, the structure of a debris disk is an important indicator of the formation and evolution processes of planetary systems, one of today's main subjects of research in astronomy.

The dust in the debris disk is heated by the radiation from the central star and re-emits stellar thermal radiation. The dust is cool, meaning it peaks in infrared (IR), while the star is hot, i.e. thermal radiation falls off in IR. Therefore, most debris disk are discovered in IR. Consequently, the first debris disks have been discovered with observations of the Infrared Astronomical Satellite (IRAS) while detecting unusually large IR emission around main sequence stars \citep{Aumann1984}. By the end of its mission, the Spitzer Space Telescope has found several hundreds of these candidate debris disks. However, there exist only some three dozen resolved images of such disks in the optical, $\rm{\beta}$ Pictoris being one of the most famous examples \citep{Smith1984}.

Due to improved stellar coronagraph techniques last two decades, a few dozen resolved debris disks have been observed at optical or near-IR wavelength, mostly with Hubble Space Telescope (HST) coronagraphy. The optical and near-IR light detected is in fact light from the central star that is scattered by the dust particles. The fraction of light reflected by the dust particles depends on their albedo $\omega$ and the optical depth $\tau$ (the latter in turn depends on the amount of dust particles). This means that measurements of the surface brightness (SB) of debris disks in optical and near-IR allow us to estimate the $\tau \omega$ (optical depth x albedo) product of the disk.  

The debris disk around the G2V star HD 107146 was the first one to be resolved in scattered light around a young solar-like star \citep{Ardila2004, Ardila2005}. Understanding this system might analogously give us clues on the early history of our own solar system or the diversity of planet formation for host stars of the similar spectral type. Therefore the system has been studied intensively in the last decade. Signs of the debris disk were first discovered by analysis of IRAS data \citep{Silverstone2000}. After this the disk has been resolved in scattered light by the Hubble Space Telescope several more times \citep{Ardila2004, Ertel2011, Schneider2014} and it has been observed in the (sub)millimeter regime by the James Clerk Maxwell Telescope \citep{Williams2004a} and by Atacama Large Millimeter/submillimeter Array (ALMA) \citep{Ricci2015}. 

HD 107146 is at a distance of approximately $27.4\pm0.4$ pc from Earth \citep{VanLeeuwen2007}. The age of the system is estimated to be between 80 and 320 Myr \citep{Williams2004a, Moor2006, Roccatagliata2009}. The ring-like disk is seen almost face-on with an inclination angle of $18.5\degr\pm2\degr$ \citep{Schneider2014} and has a celestial position angle (North-East orientated) of $58\degr\pm5\degr$ \citep{Ardila2004}. It has a high scattered light SB, i.e. it can be imaged with a high signal-to-noise ratio \citep{Ardila2004}. Models of the debris disk at sub-millimeter wavelengths show it extends from $\sim$30-150 au as seen from the central star with an approximately 8 au gap at about 80 au \citep{Ricci2015}. However, scattered light observations show an even larger disk extending to approximately $\sim$200-250 au \citep{Ardila2004,Ertel2011}\footnote{We note that sub-mm emission and scatter are sensitive to different sizes of dust grains.}.

The disk is bright in the thermal IR with $L_{IR}/L_{star} \simeq 1.2 \times 10^{-3}$, but with a visible-light scattering fraction of only $\sim7.7 \times 10^{-5}$ \citep{Schneider2014}.
The optical SB peaks at approximately 131 au suggesting more dust in the outer regions of the system \citep{Ertel2011}. Total disk masses were derived using different data sets and result in a total mass between $4.4\times10^{-7} \rm{M_{\sun}}$ and $8.5\times10^{-7} \rm{M_{\sun}}$
\citep{Ertel2011}. This makes the outer ring of HD 107146 a more extended and more massive analogue to our Kuiper Belt. The possible existence of a planet near the inner edge of the disk has been suggested to explain the double-ringed structure of the system \citep{Pearce2015}. 

Observations of HD 107146 with the HST Advanced Camera for Surveys (ACS) coronagraph in 2004 reveal a resolved background galaxy (with magnitude V=19.4, V-I=1.2) in close angular proximity to the debris disk \citep{Ardila2004}. This galaxy is called the Vermin Galaxy by \cite{Schneider2014}, due to these author's initial "annoyance" with its presence. 

These observations of the debris disk are  not contaminated by the light of the galaxy due to a large angular separation. However, due to the reflex proper motion of the HD 107146 the debris disk will transit and thus occult and dim the background galaxy presently (2017) with the periphery of the debris ring currently transiting the galaxy. This means we can use this opportunity to perform a relatively new technique to detect and characterize debris dust, called debris ring transit photometry: using a background object (in our case a galaxy) to measure attenuation by the dust in the transiting debris ring in front \citep{Zeegers2014}.

As mentioned before, if we observe a debris disk in optical light, we observe light from the central star scattered by the dust. The fraction of light which is scattered by these dust-particles depends on their albedo. However, if a background object is present, we can make use of the physical size of the dust particles to block light, independent of the albedo \citep{Zeegers2014}. Since each dust particle will block part of the light from the background galaxy, we can measure the net contribution of all debris particles in front of the galaxy. Unlike background stars, for background galaxies it is not possible to identify individual collisions between planetesimals when the debris disk stands in front of it, due to their large effective diameter \citep{Zeegers2014}. However, with the use of a background galaxy the average optical depth of the debris disk along one or more independent lines of sight can be measured. A similar method has also been used to detect dust in the case of two occulting galaxies \citep[see for example][]{Holwerda2007,Holwerda2009,Holwerda2013}.

Our goals are to determine if 
(a) the epoch 2004/2011 (on the cusp of transit) images of the Vermin galaxy could be characterized well enough to serve as baseline references for follow-on epoch transit photometry, and 
(b) there exists evidence for line-of-sight extinction (attenuation) of the galaxy light in the 2011 epoch imaging.

In this paper we will analyze relatively new scattered light images of HD 107146 observed with HST Space Telescope Imaging Spectrograph (STIS) in 2011 \citep{Schneider2014}. We will also analyze scattered light images of HD 107146 observed with the HST/ACS in 2004 \citep{Ardila2004}. In Section~\ref{s:observations} the 2004 and 2011 observations are summarized. Then these data are used to create a model of the Vermin Galaxy in Section~\ref{s:IMFIT}. Using this model we can perform debris ring transit photometry explained in Section~\ref{s:debrisringtransit}. Section~\ref{s:results} will present the results. They are compared with previous research in Section~\ref{s:comparison}. The results are discussed in Section~\ref{s:discussion}. Conclusions are given in Section~\ref{s:conclusions}. At last possibilities for future research are discussed in Section~\ref{s:future}.

\section{HST observations from 2004 (ACS/HRC) and 2011 (STIS)}
\label{s:observations}

\begin{figure*}
	\centering
    \includegraphics[width = 2\columnwidth]{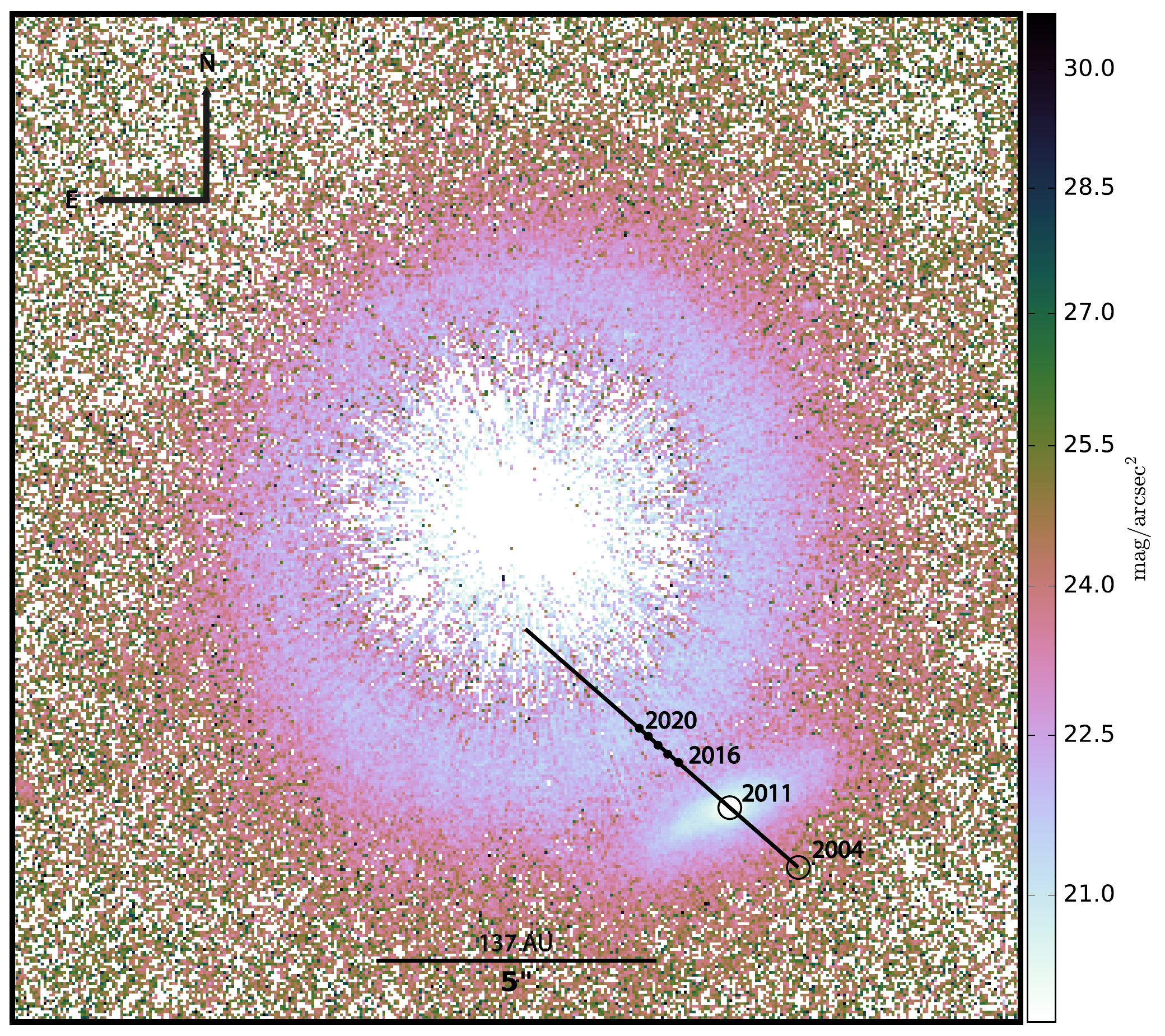}
    \caption{Image of HD 107146 taken in 2011 with HST/STIS PSF subtracted coronagraphy. We analyze these data to find attenuation of the galaxy by dust particles in the debris disk. The black line indicates the reflex proper motion of the galaxy in the upcoming years calculated from the proper motion of HD 107146 by~\protect\cite{VanLeeuwen2007}. The center of the Vermin Galaxy will reach the brightest part of the disk in 2020 \citep{Schneider2014}. The bar indicating the length of 5\arcsec shows the distance in au assuming a face-on orientation of the disk.}
	\label{fig:data2011}
\end{figure*}

In this research we use two HST data sets: (1) data of the debris disk around HD 107146 from 2004 discussed by \citep{Ardila2004}. This data set contains two reduced images taken with the ACS High-Resolution Channel in the Broad I- and the V-band. In these images the Vermin galaxy is still out of transit (see the positions marked in Figure~\ref{fig:data2011}). (2) one relatively new STIS 50CCD image of the debris disk around HD 107146 from 2011, which is also taken in the optical close to the ACS/HRC V-band central wavelength, but with a much broader bandwidth \citep{Schneider2014}. In this image the Vermin galaxy is already at the edge of the debris disk (see Figure~\ref{fig:data2011}). Observational parameters of the three images are summarized in Table~\ref{tab:observationalparameters}. Both data sets are discussed in more detail below.

\begin{table}
	\centering
	\begin{tabular}{l c c c}
{\textbf{Images}} & {} & {} & {}\\ \hline
{Year (UT)} & {2004} & {2004} & {2011} \\
{Instrument} & {ACS/HRC} & {ACS/HRC} & {STIS} \\
{Passband} & {F606W (V)} & {F814W (I)} & {50CCD} \\
{Central wavelength [$\angstrom$]} & 
{5907} & {8333} & {5850} \\
{FWHM [$\angstrom$]} 
& {2342} & {2511} & {4410} \\
{Total integrated time [s]} & {2330} & {2150} & {9862} \\
\hline \\
    \end{tabular}
  	\caption{Observational parameters of our data sets. Note the two 2004 images are observed in two different filters: the V- and the I-band. The 2011 image has been observed without a filter, therefore the central wavelength and full width at half maximum (FWHM) are the STIS instrumental parameters. The total integrated time indicates the total exposure time combining all raw data, therefore, the actual exposure time is shorter in the 2011 case.}
  	\label{tab:observationalparameters}
\end{table}

\subsection{The 2004 ACS/HRC data}
\label{s:data:2004}

HD 107146 was observed with ACS/HRC on Universal Time (UT) June 5th, 2004. The reference star HD 120066, used to obtain the point spread function (PSF), was observed with the ACS/HRC on UT July 20th, 2004. Two filters were used: the F606W (V-band) and F814W (I-band) filters. A short direct exposure was followed by a coronagraphic exposure using a 1.8$\arcsec$ diameter mask. The observational parameters of these exposures are summarized in Table \ref{tab:observationalparameters}.

The direct HD 101746 image was reduced in seven steps: 
(1) the flux of the direct image was integrated within a circular aperture with radius $>$ 6 $\arcsec$, including the saturated stellar core, to determine the magnitude of the target and PSF reference star 
(2) counts/second were transformed to magnitudes using the STSDAS synthetic photometry package SYNPHOT, 
(3) the aligned, flux-scaled coronographic image of HD 120066 was subtracted from the coronographic image of HD 107146, 
(4) the resulting image was smoothed by a 3x3 median filter, 
(5) the image was corrected for geometrical distortion, 
(6) smoothed again with a 5x5 median filter, and 
(7) re-binned by a factor of two to improve the signal-to-noise ratio, resulting in 0\farcs127 and 0\farcs173 spatial resolution in F606W and F814W respectively  \citep{Ardila2004,Ardila2005}.

\subsection{The 2011 STIS data}
\label{s:data:2011}

We use HST/STIS 2011 coronagraph observations of the debris disk around HD 107146. The HD 107146 system and the PSF reference star HD 120066 were contemporaneously observed together at two epochs: UT February 22th, 2011 and UT May 30th, 2011. A F25ND3 filter was used for the target acquisition imaging to place the star behind the corongraphic mask. The coronagraphic images at two occulting mask positions were both unfiltered in the broad bandpass of the STIS 50CCD optical channel. Initially, the idea was to create a six-roll combined analysis quality image, but eventually a five-roll combined analysis quality image was made in order to improve prevent degradation due to non-optimal data at one roll angle. The resulting image is represented in Figure~\ref{fig:data2011}. 

The field of view (FOV) of this image is $400\times400$ pixel, which corresponds with $20.3\rm{\arcsec}\times20.3 \rm{\arcsec}$. The debris disk has not been de-projected to face-on. The effective inner working angle ($IWA$) along the major axis (south-east to north-west) of the inner region of the debris disk is $IWA_{\rm{eff}} = 0.41\rm{\arcsec}$. The image scale is $0.05077\rm{\arcsec/pixel}$ and $1 \ \rm{count/s} = 4.55\times10^{-7} \rm{Jy}$ (with GAIN = 4), which corresponds to $1 \ \rm{count/s/pixel} = 0.1765 \ \rm{mJy/\arcsec}^2$.

Compared with the observations by \cite{Ardila2004}, the galaxy is closer towards the debris disk up to 6.25$\rm{\arcsec}$ from the star, due to the proper motion of the central star. The center of the Vermin Galaxy will be behind the brightest part of the debris disk  (which is co-moving with the star) around 2020 \citep{Schneider2014}. The 2011 image has improved sensitivity to trace dust depletion in the inner regions to $\sim$60 au compared to previous ACS and Near Infrared Camera and Multi-Object Spectrometer data. The debris disk has a STIS 50CCD band integrated flux density of $F_{\rm{disk}} = 0.40 \ \rm{mJy}$ (integrated over the model starting at 0\farcs4 or $\sim$55 au) and a visible light scattering fraction of $7.7 \times 10^{-5} \pm 4.0 \times 10^{-6}$. This is interesting in context of its 16 times larger IR excess. From previous ACS data dust depletion in the interior regions were inferred \cite{Ardila2004}, but the extent of this depletion was uncertain. In the new 2011 data this decrement in the SB can clearly be seen to the contrast-limited angular distance of $r\sim2"$.
For a more complete overview on the observation and data set we refer the reader to the original paper by \cite{Schneider2014}.

\section{Modeling the Vermin Galaxy with IMFIT}
\label{s:IMFIT}

We need to determine the difference between the brightness of the non-occulted (epoch 2004) and partly occulted Vermin Galaxy (epoch 2011), and we require a model of the Vermin Galaxy since this model will represent the amount of light we expect before the occultation by the debris ring.

Modeling the galaxy means finding a two-dimensional light-profile that fits the observed light profile of the Vermin galaxy as well as possible. Different galaxy types have their own specific light-profiles and hence, if we can find a fitting model for our galaxy, this not only tells us something about its parameters, but also about the type of galaxy. In order to find the right model for the Vermin galaxy, we start by applying simple models to describe its light profile.

To model the Vermin galaxy we looked at two galaxy modeling packages - {\sc imfit} \citep{Erwin2015a} and {\sc galfit} \citep{Peng2002}. We tried both programs, but use {\sc imfit} because of its stability. In order to create a model, a PSF of the instrument is needed. The PSFs of the HST/ACS and HST/STIS are both created with {\sc TinyTim}\footnote{\url{http://www.stsci.edu/hst/observatory/focus/TinyTim}} online software \citep{Krist2011}.

{\sc imfit} uses $\chi^2$-optimization to find the best fitting model. To do this optimization three input images are used: (1) the observed image, (2) a mask image indicating the location of bad pixels, and (3) a root-mean-square-map of the image. Such an image can be created with {\sc Sextractor} \citep{Bertin1996}. From these images the value of the weight $w_i$ of each pixel is calculated. The weight is defined as $w_i = z_i / \sigma_{I,i}^2$, where $z_i = 1$ for unmasked pixels, $z_i = 0$ for masked pixels and $\sigma_{I,i}$ the root-mean-square of each pixel. Subsequently, the $\chi^2$-value, defined as: 
\noindent
\begin{equation}
	\chi^2 = \sum_{i} w_i (I_{d,i}-I_{m,i})\,,
\end{equation}
\noindent where $I_{d,i}$ is the data pixel value and $I_{m,i}$ is the model pixel value, is optimized. In this context optimizing means minimizing $\chi^2$. Every iteration the model is changed by translating, geometrical scaling, rotating, and adjusting parameters of 2D-functions, reconstructing the light profile in the image.

In order to model the Vermin galaxy properly, we first analyze the out-of-transit observations by the HST/ACS coronagraph in 2004 \citep{Ardila2004}. We assume that the light of the galaxy is still unaffected by the debris disk, due to their large angular separation. We chose to separately model the two 2004 images taken with a broad V- and I-filter to check if they are consistent. Since the 2004 and 2011 observations are taken in different spectral filters with different central wavelengths (see Table~\ref{tab:observationalparameters}), we can not use them for a direct photometric comparison, but we can use them to constrain model parameters: the shape, orientation and size of the galaxy. These parameters then can be used as guide to model the Vermin galaxy at the edge of the debris disk in the 2011 data. Additionally, in the context for the epoch 2017 - 2019 STIS transit observations, the 2011 galaxy ("pre-transit") image will serve as a reference epoch template for differential photometric measurements.

\subsection{Out of transit observation: 2004 HST/ACS}
\label{sec:imfit:acs}

We model the galaxy with a S\'{e}rsic model \citep{Erwin2015a}. This S\'{e}rsic model is generally used to fit the light-profile of main galaxy components (disks and bulges). The model is defined as: 
\begin{equation}
	{I(r) = I_{\rm{e}} \exp{ \bigg( -b_n \Big( \big(\frac{r}{r_{\rm{e}}} \big)^{1/n} -1\Big) \bigg) }}\,,
\end{equation}
\noindent where $I(r)$ is the surface brightness (SB) at radius $r$ from the center, $I_{\rm{e}}$ the SB at the half-light radius $r_{\rm{e}}$ and $n$ the index controlling the shape of the intensity profile. The value of $b_n$ is formally the solution of the transcendental equation, however in {\sc imfit} it is calculated with a polynomial approximation \citep{Erwin2015a}.

A special case of this model that can be used is the exponential disk model:
\begin{equation}
	{I(r) = I_{\rm{0}}  \exp{\Big(\frac{-r}{h} \Big)}}\,,
\end{equation}
\noindent where $I_{\rm{0}}$ is the central SB and $h$ the scale length. This model is empirically known to be a good default for the light-profile of the disk of spiral galaxies.

We use {\sc imfit} to optimize the parameters for these two models on the Vermin galaxy. By subtracting the modeled and PSF-convolved light-profile from the observed light profile of the galaxy, we get the residual light-profile, which we can analyze for substructures intrinsically present in the Vermin galaxy. Analysis of the 2004 residuals shows a single exponential disk model (equation 3) works well for our purpose in the outer regions of the galaxy, since the pixel values in the residual image at the original position of the galaxy are indistinguishable from the background noise. However, the inner regions of the galaxy show residuals that require an additional structural component. The two-component models that we have applied to the Vermin galaxy are: combinations of two S\'ersic models, two exponential models, an exponential disk with a S\'ersic bulge and the latter with the addition of a Gaussian ring. 
The optimal two-component model is a S\'{e}rsic bulge and an exponential disk. The best-fitting parameters and the corresponding uncertainties for this model found with {\sc imfit} are listed in Table~\ref{tab:models}.

\begin{table*}
	\centering
    \begin{tabular}{  l  c  c  c  }
      \hline 
      {\textbf{Dataset}} & {\textbf{2004}} & {\textbf{2004}} & {\textbf{2011}}\\
      {\textbf{}} & {\textbf{V} (ACS/F606W)} & {\textbf{I} (ACS/F814W)} & {\textbf{Clear} (STIS/50CCD)}\\
      {\textbf{pixel scale}} & {0\farcs025/pixel} & {0\farcs025/pixel} & {0\farcs05077/pixel } \\ 

      \hline \hline \\
      {\textbf{General}} & {} & {} & {} \\ \hline
      {reduced $\chi^2$} & {2.11} & {1.31} & {2.60} \\ 
      {$x_{\rm{0}}$ [pixel]} & {837.12 $\pm$ 0.03} & {836.87 $\pm$ 0.03} & {922.97 $\pm$ 0.01} \\ 
      {$y_{\rm{0}}$ [pixel]} & {321.85 $\pm$ 0.03} & {322.50 $\pm$ 0.02} & {653.99 $\pm$ 0.01} \\
\hline  \\
      
      {\textbf{Disk (Exponential)}} & {} & {} & {} \\ \hline
      {position angle [$\degr$]} & {-66.23 $\pm$ 0.19} & {-66.25 $\pm$ 0.17} & {-63.26 $\pm$ 0.15} \\ 
      {ellipticity} & {0.653 $\pm$ 0.003} & {0.623 $\pm$ 0.004} & {0.625 $\pm$ 0.003} \\ 
      {$I_{\rm{0}}$ [{counts/second/pixel}]} & {0.119 $\pm$ 0.001} & {0.150 $\pm$ 0.002} & {0.129 $\pm$ 0.001} \\ 
      {h [pixel]}$^*$ & {40.44 $\pm$ 0.30} & {33.70 $\pm$ 0.44} & {20.77 $\pm$ 0.20} \\
      \hline  \\
      
      {\textbf{Bulge (S\'ersic)}} & {} & {} & {} \\ \hline
      {position angle [$\degr$]} & {-58.28 $\pm$ 1.45} & {-30.57 $\pm$ 2.71} & {-74.92 $\pm$ 0.72} \\ 
      {ellipticity} & {0.183 $\pm$ 0.008} & {0.140 $\pm$ 0.010} & {0.284 $\pm$ 0.006} \\ 
      {$n$} & {0.84 $\pm$ 0.01} & {1.00 $\pm$ 0.02} & {0.80 $\pm$ 0.01} \\ 
      {$I_{\rm{e}}$ [{counts/second/pixel}]} & {0.065 $\pm$ 0.001} & {0.053 $\pm$ 0.001} & {0.094 $\pm$ 0.001} \\
      {$r_{\rm{e}}$ [pixel]}$^*$  & {8.98 $\pm$ 0.09} & {8.93 $\pm$ 0.22} & {4.53 $\pm$ 0.06} \\
      \hline  \\
      \multicolumn{4}{l}{$^*$note the apparent difference between 2004 and 2011 in scale length and effective radius} \\
      \multicolumn{4}{l}{are also due to the difference in plate scale between ACS and STIS of respectively}\\
      \multicolumn{4}{l}{$\sim$0.028 $\times$ 0.025$\arcsec$/pixel (north-south x west-east) and 0.05077$\arcsec$/pixel.}\\
    \end{tabular}
  \caption{Best fit parameters for the 2004 images in the V- and I-band and for the 2011 image. The position angle is measured counter-clockwise from the positive y-axis (i.e. north-east orientated) in degrees. Ellipticity is defined as $1 - b/a$ where $b$ is the semi-major axis and $a$ the semi-minor axis. The values of the central position of the galaxy $x_{\rm{0}}$, $y_{\rm{0}}$, the scale length $h$ and effective radius $r_{\rm{e}}$ as defined in Section~\ref{s:IMFIT} are measured in pixel and the central SB $I_{\rm{0}}$ and SB at the effective radius $I_{\rm{e}}$ are measured in counts/second/pixel. We find that $n \leq 2$, which is an indication that we are dealing with a pseudo-bulge galaxy \citep{Fisher2008}. The apparent difference between the position angles of the bulge component of each model can be understood by low ellipticity of the bulge models; these bulges are close to spherical and thus rotationally symmetric giving the Position angle of the "major axis" a random value with a low formal error. The uncertainties are generated by the fitting program {\sc imfit} with the use of a background root-mean-square image created with {\sc SEXtractor} \citep{Bertin1996} from the original image. These uncertainties are purely mathematical \citep{Erwin2015a} and should be regarded as a lower limit on the real errors.}
\label{tab:models}
\end{table*}

\begin{figure*}
	\centering
    \includegraphics[width=2 \columnwidth]{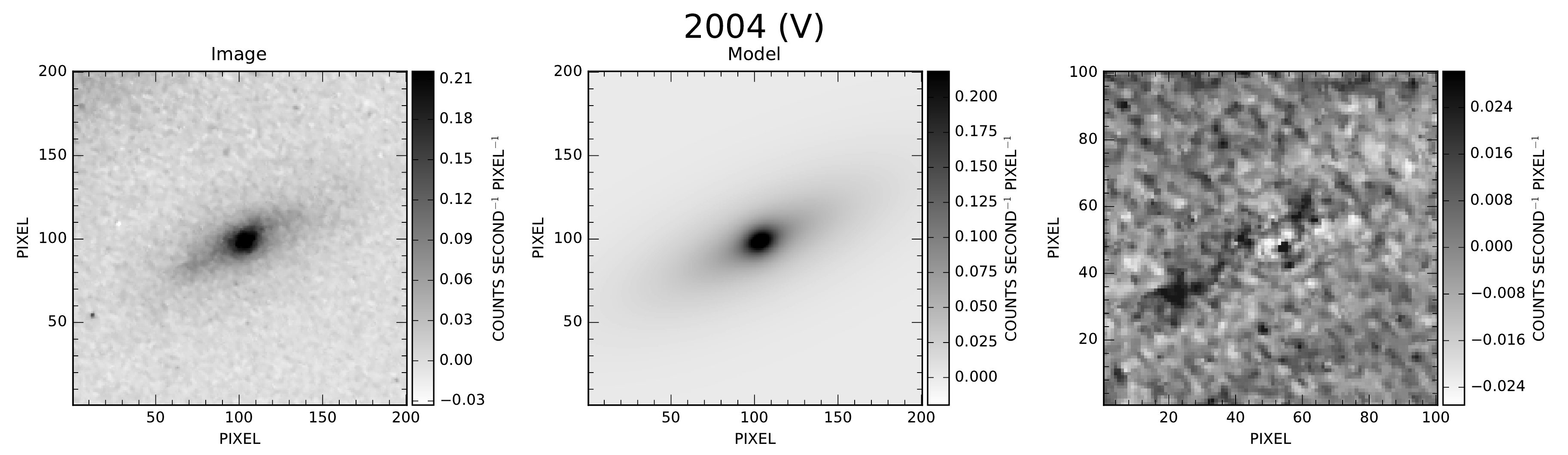}
    \includegraphics[width=2 \columnwidth]{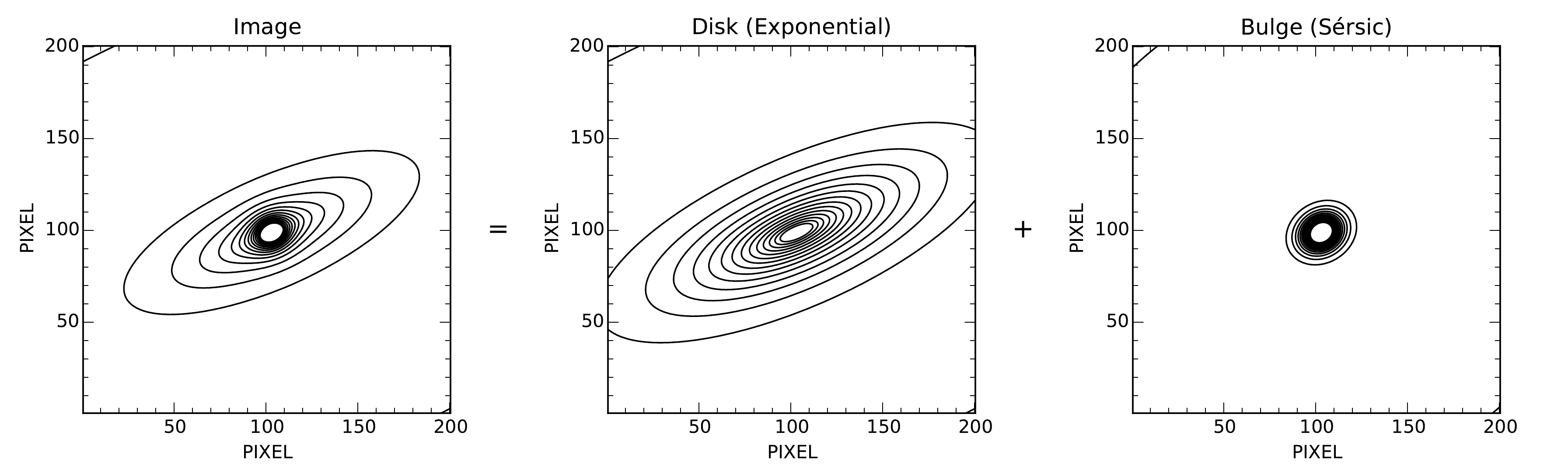}
    \\
    \includegraphics[width=2 \columnwidth]{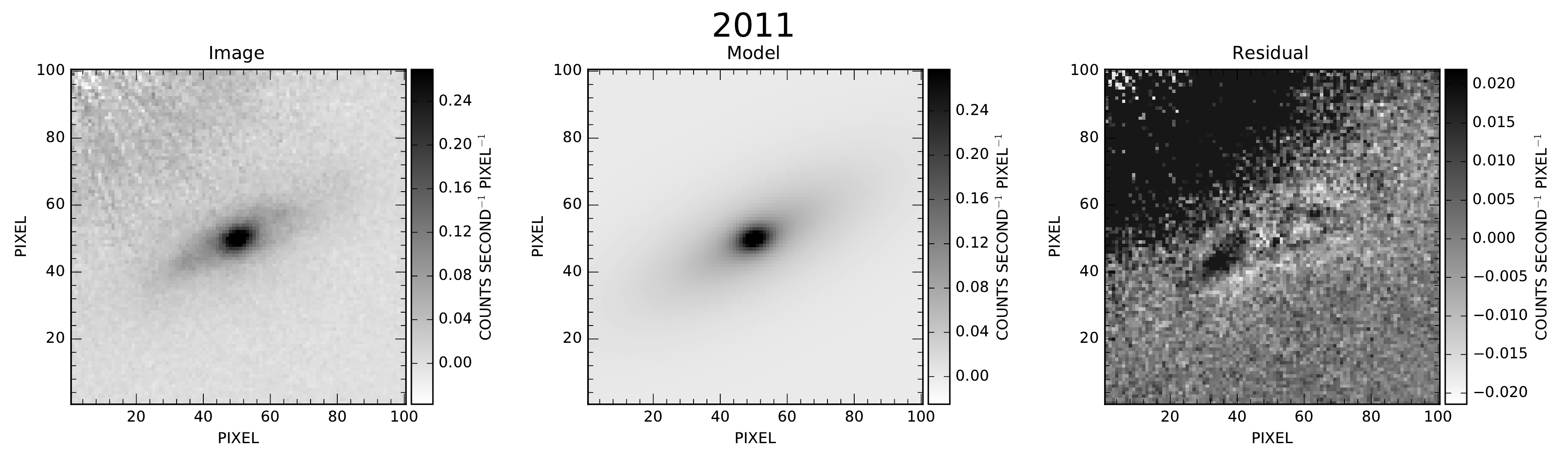}
    \includegraphics[width=2 \columnwidth]{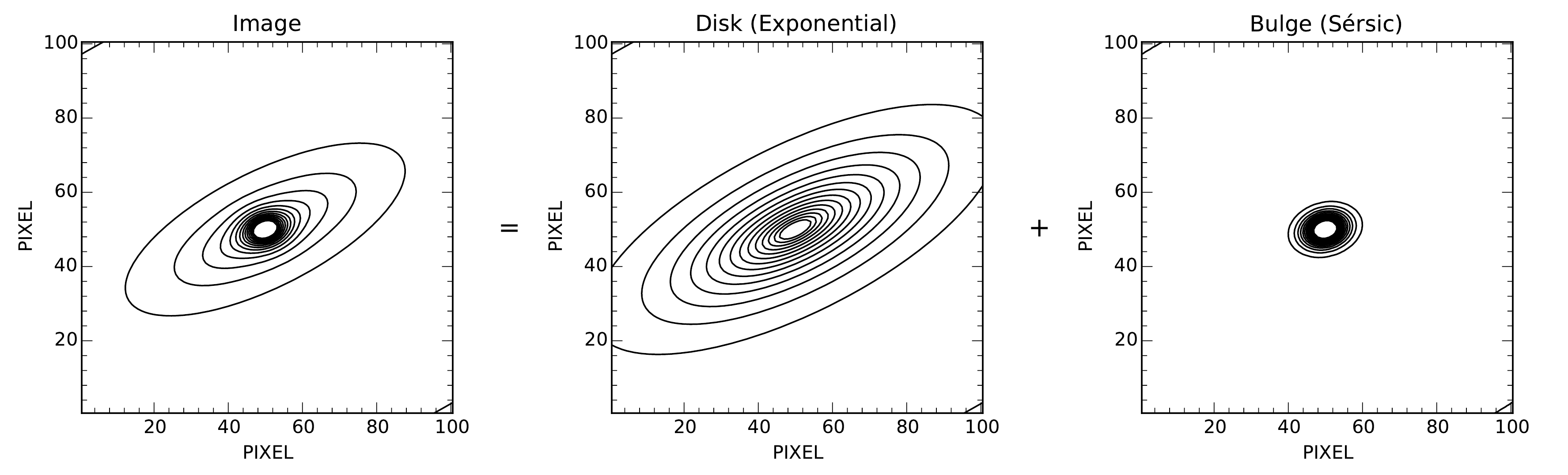}
    \caption{\textit{First row:} the 2004 ACS F606W wide V-band image of the Vermin Galaxy, the best model of the galaxy and the residual image. This figures look very similar in the ACS F814W wide I-band. \textit{Second row:} the 2004 V-band model of the Vermin galaxy and its components. The model consists of the sum of two functions: an exponential disk and a S\'ersic (pseudo-)bulge. The contours are linearly-spaced levels of equal SB, although the levels differ between the three figures. The model and its components look very similar in the ACS F814 wide I-band. \textit{Third row:} the 2011 image of the Vermin Galaxy, the best model of the galaxy and the residual image. The debris disk can already be seen in the left-corner of the image. \textit{Fourth row:} the 2011 model of the Vermin galaxy and its components. Like for the 2004 model, the model consists of the sum of two functions: an exponential disk and a S\'ersic (pseudo-)bulge. The contours are linearly-spaced levels of equal SB, although the levels differ between the three figures.}
    \label{fig:imfit}
\end{figure*}

The original 2004 image data (\S \ref{s:data:2004}), the best-fitting model of the galaxy, the residual and the components of the model are shown in the upper two rows of Figure~\ref{fig:imfit}. Here we only show the V-band image, because the V-band and I-band figures are almost identical. 

Because the Vermin galaxy can best be modeled with an exponential disk with a S\'{e}rsic bulge (least amount of visible residual and lowest $\chi^2$), we can conclude that it is a distant spiral galaxy. 
The best-fitting S\'{e}rsic profile of the bulge has an index of $n \leq 2$ (Table~\ref{tab:models});  the Vermin galaxy has a pseudo-bulge as opposed to a ``classical" bulge ($n \sim 4$). Pseudo-bulges are systematically flatter than classical bulges, thus more disk-like in both their morphology and shape and also quite common \citep{Fisher2008}. The relative lack of prominence of the bulge indicates that the galaxy is a late-type galaxy (Sc). 

The upper two rows of Figure~\ref{fig:imfit} show the 2004 residual is close to the background noise in all parts of the galaxy. The model is sufficient to describe the light profile of the galaxy; two components suffice and the remaining structure is of the scale of the noise in the image.
The Vermin galaxy is therefore an excellent candidate for performing debris ring transit photometry. The smoothness of the galaxy also indicates that the Vermin galaxy is relatively distant, i.e., such that detailed structures cannot be resolved (e.g. clear spiral arms or bars). This serves our purposes well, since the galaxy can be modeled well by a simple axis-symmetrical model, even at the bluer wavelengths which typically show more substructure (thanks to star-formation). 

\begin{figure}
	\centering
    \includegraphics[width = \columnwidth]{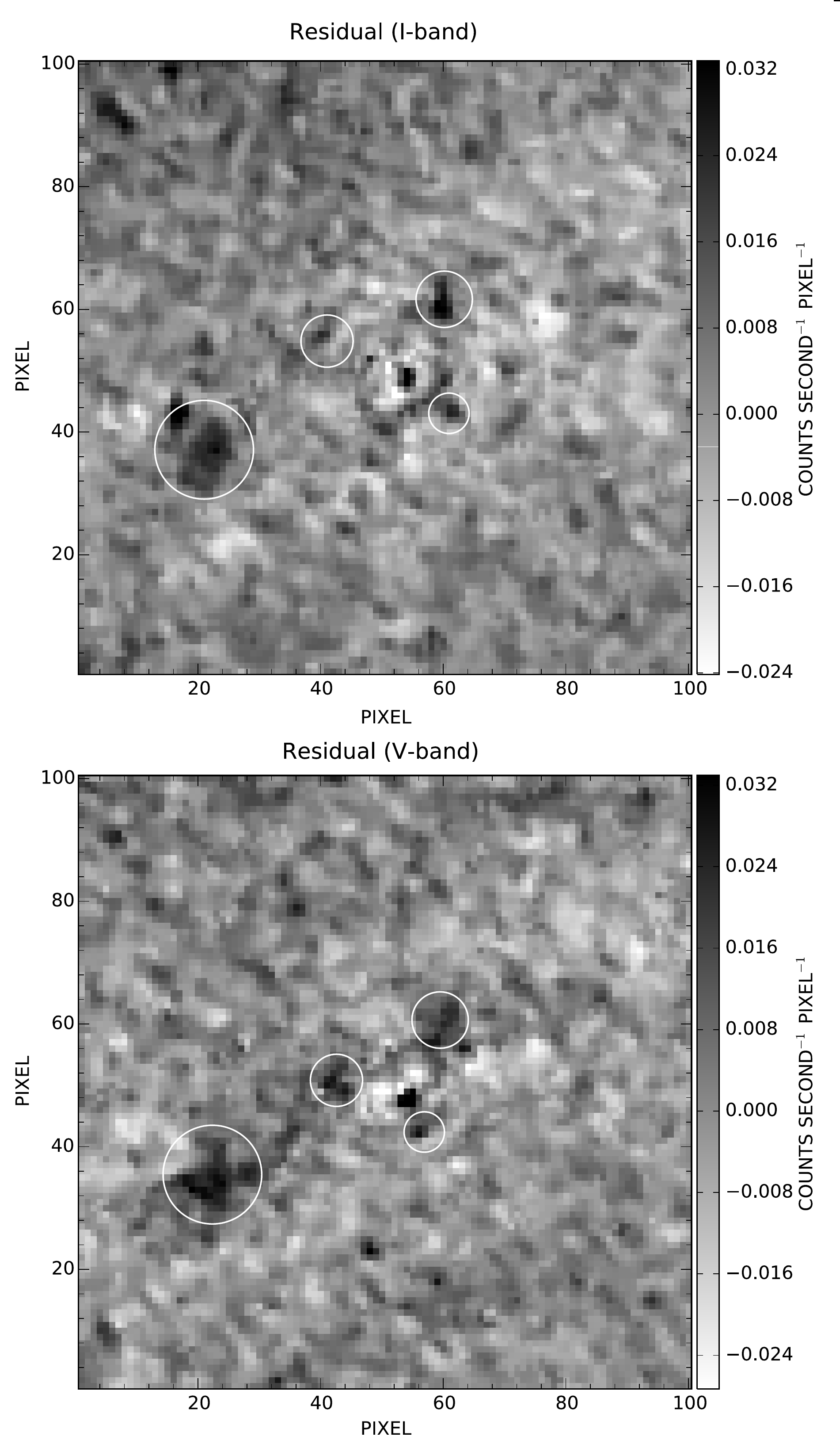}
    \caption{F606W wide V-band (top) and F814W wide I-band (bottom) images after subtracting the best-fit galaxy model. If the model underestimates the amount of light the regions appear darker. These dark overdensity regions appear in both spectral bands. We believe these to be star-formation regions. This is supported by the fact that the regions appear at almost the same locations in both images, suggesting they are intrinsically present in the Vermin galaxy. The apparent change in the positions of the overdensities relative to the center of the galaxy can be explained by the uncertainty in the central position of the galaxy.}
    \label{fig:residuals2004}
\end{figure}

However, we can still see some remnant structures in the residual image. These most-probably are small-scale star-formation regions and therefore are brighter then the model predicts. These regions appear almost on the same locations in the V-band and I-band, however they differ a small amount in size and brightness (see Figure~\ref{fig:residuals2004}). These regions will, of course, persist on transit timescales and so are expected to be in all subsequent transit epoch images.
Since the 2004 FITS-files do not contain sufficient photometric information (bias subtracted and effective gain) we are unable to calculate the color difference (V-I) between the two images.

Luckily there are few overdensities in the residuals and they are restricted to the central region ( within an exponential scale-length $r<h$). Nonetheless, it is important to keep in mind there are some overly bright regions, when we want to look for attenuation by dust in the transiting disk in the 2011 image and future epochs.

\subsection{On the edge of transit observation: 2011 HST/STIS}
\label{sec:imfit:stis}

We start with the galaxy's size (effective radius), orientation (position angle), and shape (profile choices of a Sersic Bulge and Exponential disk) at the values we found for our 2004 model with the galaxy fully unobscured by the debris ring. We can then model the Vermin galaxy at the periphery of the debris ring in the 2011 observations with the use of {\sc imfit} again. This leaves relative flux contributions, axis ratio, central position, profile shape (Sersic Index), and total flux as free parameters in our {\sc imfit} model. We must also take into account the difference in the STIS 50CCD passband compared to ACS "wide" V and I (see Table~\ref{tab:observationalparameters}). 

We know that the 2011 observation is taken in a wide wavelength range from the near-IR up to the UV (see STIS handbook Figure 5.1), while the 2004 observations are taken in two much narrower ranges centered in the optical, well within the STIS 50CCD passband. From this we expect for the 2011 model a similar light-profile as the 2004 models, however, perhaps with more substructure. This means we cannot simply scale and overlay the 2004 model onto the 2011 image. However, we can still make use of the best-fitting parameters inferred from the 2004 model. Firstly, we can use the 2004 model parameters as initial guess in {\sc imfit}. Secondly, we can use it as a sanity check. Thirdly, we can compare the 2004 and 2011 models in more detail.

Since the STIS 50CCD band central wavelength lies closer to the V-filter central wavelength than the I-filter central wavelength, we take the best fit parameters of the Vermin galaxy in the 2004 V-filter as the starting point of our fit for the 2011 model. From the models of the 2004 observations, we already know that the galaxy can be modeled well by an exponential disk with a S\'{e}rsic (pseudo-)bulge. We expect the same light-profile for the 2011 model and keep this choice of morphological parameterizations. 

We will start by fitting just an single exponential disk to the galaxy for the 2011 observation. Here the position angle, ellipticity and the new scale length $h$ (where the scale length is adapted for the different plate scales of the two instruments, see Tables \ref{tab:observationalparameters} and \ref{tab:models}) will be fixed for our initial guess, because these values are expected to be similar to the 2004 values. We then try to find the best fit for the galaxy with {\sc imfit}, by adding a (pseudo-)bulge. 

\begin{figure}
	\centering
	\includegraphics[width=\columnwidth]{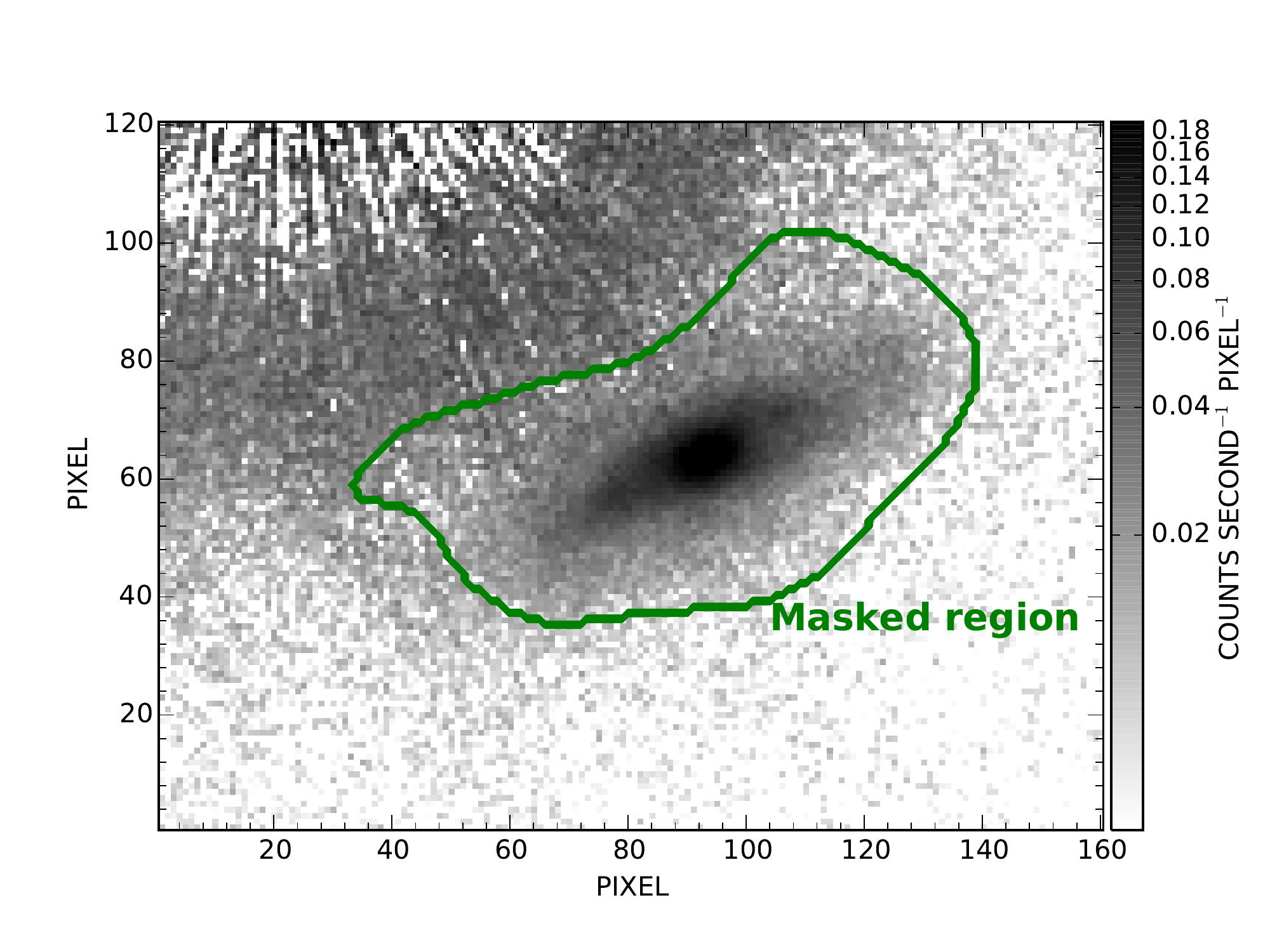}
	\caption{Image of the 2011 Vermin Galaxy. The green line shows the border of the masked region: all pixels inside this region are used during the fitting procedure and all pixels outside this region are not taken into account. This shape allows for masking the brightest part of the debris disk efficiently and has been created with {\sc Sextractor}.}
	\label{fig:mask}
\end{figure}

During this procedure we have to take into account light from the close angular proximity debris disk that could affect the fitting-procedure at the Northern side of the vermin galaxy's edge. Therefore this polluted region is masked during the fitting procedure. This mask is made with {\sc sextractor}  \citep{Bertin1996} by finding the approximate size of the galaxy and masking everything outside this region. The exact masked region is shown in Figure~\ref{fig:mask}. 

It turns out that the best fitting model for the galaxy consists (as expected) of an exponential disk and an additional S\'{e}rsic (pseudo-)bulge with parameters quite similar to those found in 2004. The best fitting parameters for the Vermin galaxy for the 2011 observations are shown in Table~\ref{tab:models}. The 2011 image, best-model of the galaxy, residual and components of the model are shown in the lower two rows of Figure~\ref{fig:imfit}. In this final fit the 2004 model for the V-band supplied the initial guess for the 2011 model. 

\subsection{Comparing the 2004 and the 2011 models}

The best-fitting models of the Vermin Galaxy in each epoch consists of an exponential disk with a S\'{e}rsic (pseudo-)bulge, i.e. a spiral galaxy. All models seem to have similar parameter values when corrected for the difference in the plate scale. This is also indicated by the almost equal bulge-to-total flux ratios of the galaxy models: $B/T$: ${B/T}_{\rm{V,2004}}$ = 0.100, ${B/T}_{\rm{I,2004}}$ = 0.097 and ${B/T}_{\rm{2011}}$ = 0.102. These results are expected, since the 2004 I- and (especially) V-band wavelength regimes overlap mostly with the 2011 STIS wavelength regime and therefore we do expect a similar appearance of the galaxy for these three images. However, when looking at the residuals in Figure~\ref{fig:imfit}, one can see that more residual structure is present in the 2011 residual compared to the 2004 residual: the 2004 image seems closer to the background noise than the 2011 image. This is probably due to the fact that the 2011 image has a  broader wavelength range and because the STIS CCD is more sensitive at bluer wavelengths, or that the ACS images were smoothed slightly by \cite{Ardila2004} or a combination of the above. Therefore, one would expect to see more substructure in the 2011 case. 

In conclusion, the models are overall consistent in the outer regions and a two-component (disk+bulge) model is sufficient to describe the flux of the galaxy in all three observed bands. The remaining over/underdensities are constrained to the inner parts of the galaxy ($r<h$) and on small scales. 
If one averages over the whole or large sections of the galaxy, the {\sc IMFIT} model will describe  the overall flux with reasonable accuracy.

\begin{figure*}
	\centering
    \includegraphics[width = 2\columnwidth]{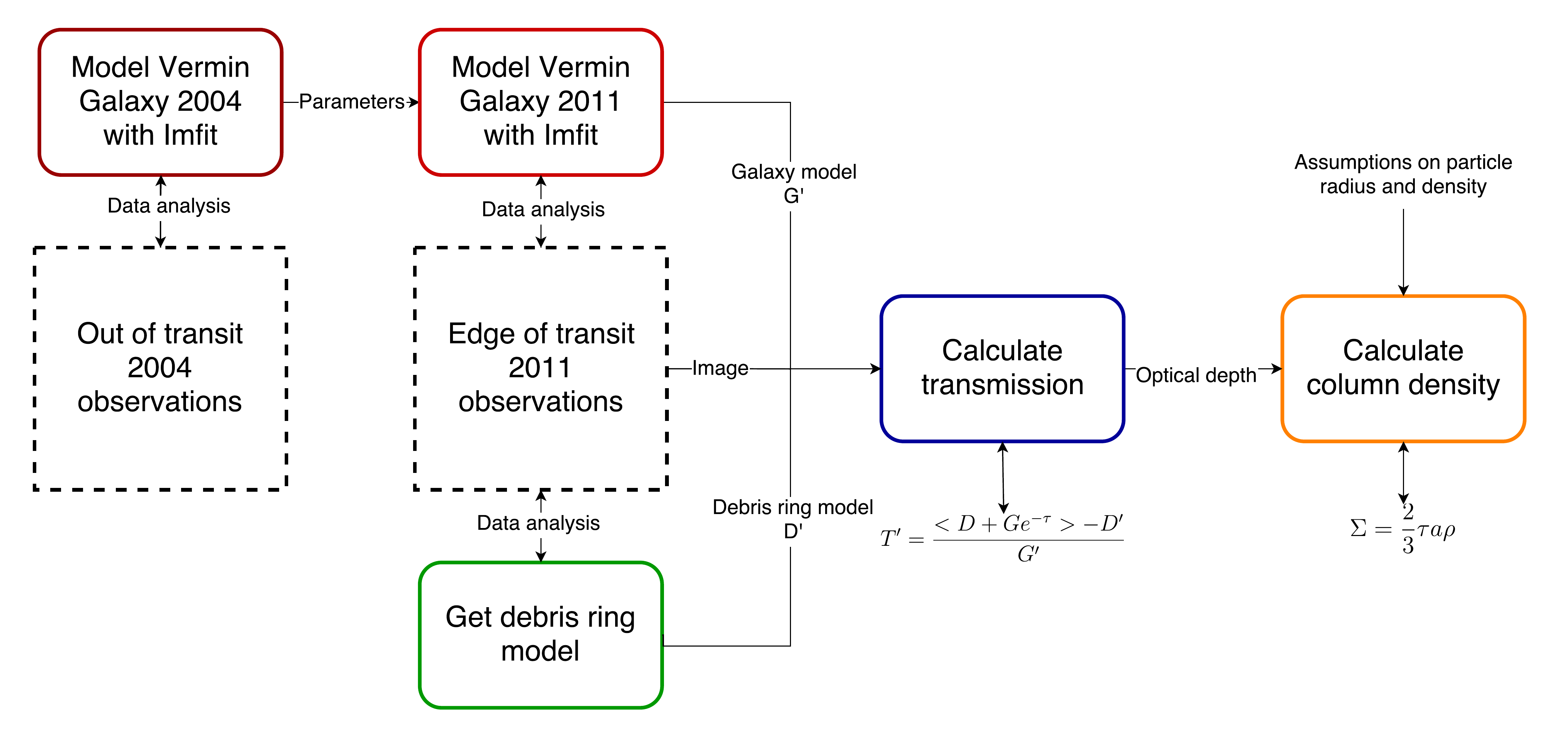}
    \caption{A flowchart overview of the debris ring transit photometry method.  First, a galaxy- and a debris ring model are obtained. Secondly, these are combined with the image to determine the transmission. Thirdly, the column density is calculated. The complete process is described in more detail in Section~\ref{s:debrisringtransit}. Note that the debris ring modeling was done by \protect\cite{Schneider2016} and the reader is referred to their paper for the details of the modeling process.}
    \label{fig:flowchart}
\end{figure*}

\begin{figure*}
	\centering
    \includegraphics[width = 2\columnwidth]{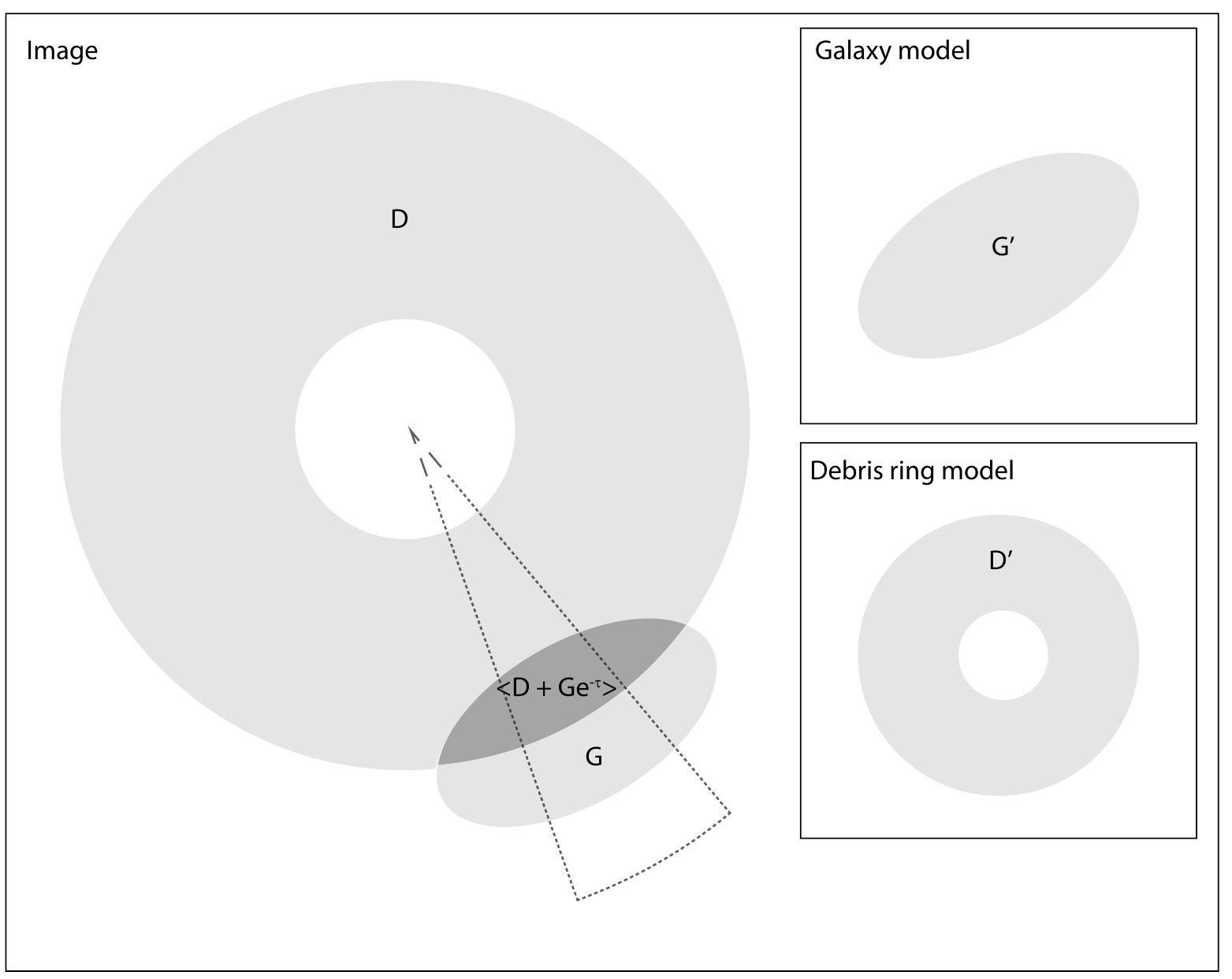}
    \caption{A schematic overview of how we determined the line-of-sight transmission profile through the disk. First note that we simplified the figure by presenting a circularly symmetric disk, but in fact we are dealing with an (almost-circular) elliptical shaped debris disk with a Henyey-Greenstein-like scattering phase angle brightness asymmetry. The radial transmission profile is determined in the elliptical region centered on the vermin galaxy with $\pm 5 \degr$ on either side (so with a total extension of $10 \degr$). $D$ denotes the disk, $G$ the Vermin galaxy and $<D+e^{-\tau}G>$ the region where they overlap. The primed values denote the models we used. $D'$ denotes the debris ring model and $G'$ prime denotes the model of the galaxy.}
    \label{fig:overview}
\end{figure*}
 
\section{Debris ring transit photometry}
\label{s:debrisringtransit}

The best-fitting model for the Vermin Galaxy can serve as a close proxy for the 2011 epoch in future observations and the column density of the debris disk superimposed along the line-of-sight to the galaxy can be determined using debris ring transit photometry. We also require a model of the debris ring around HD 107146. Here we shall use the debris ring model of \cite{Schneider2016} and the reader is referred to this paper for the details of the modeling process. In this case the Vermin Galaxy is the background object that is being transited by the debris ring, which we can use for the photometry. This method is described in this section and a flowchart overview of this method can be seen in Figure~\ref{fig:flowchart}.

\subsection{Determining the radial transmission profile from the image}
\label{s:radialprofile}
The transmission 
through the disk along the line-of-sight to the galaxy is defined as follows:
\begin{equation}
	T = e^{-\tau} = \frac{\Phi^i_e}{\Phi^t_e}\,,
    \label{eq:deftransmission}
\end{equation}
\noindent where $T$ is the transmission, $\tau$ the optical depth, $\Phi^{\rm{i}}_{\rm{e}}$ the radiant flux (from the galaxy) received by the debris disk and $\Phi^{\rm{t}}_{\rm{e}}$ the flux transmitted by the debris disk. If there is no attenuation at all then $T$=1, since the radiant flux received and radiant flux transmitted are equal. In the case we do have attenuation by dust the radiant flux transmitted will be smaller than the radiant flux received and therefore $0 < T < 1$.

To clarify how we calculated a stellocentric radial transmission profile, we will first state some definitions about how we name the different components in the image: $D$ is the real (observed) SB of the debris disk, $G$ the real SB of the Vermin Galaxy, $D'$ the debris disk template and $G'$ the model of the Vermin galaxy (see Figure~\ref{fig:overview}). Using these definitions we can express the transmission as
\begin{equation}
	T' = \frac{<D+Ge^{-\tau}> - D'}{G'}\,,
	\label{eq:transmission}
\end{equation}
\noindent where the prime in $T'$ denotes that we are calculating the transmission under the assumption $D'$ and $G'$ are a perfect representation of $D$ and $G$. We will discuss the validity of these assumptions in Section~\ref{s:discussion}. Figure \ref{fig:overview} shows a schematic overview of the different components. Using Equation~\ref{eq:deftransmission} we will calculate the radial transmission profile in the elliptical region containing the galaxy as is shown in Figure \ref{fig:overview}.

\subsection{\label{s:columndensity}Determining the column density from the transmission}

In addition to measuring the attenuation of light by the dust, the debris ring transit measurements allow us to estimate the amount of dust responsible for the attenuation. In order to determine this we calculate the column density from the radial transmission profile. We do this by first using 
Equation~\ref{eq:deftransmission} to calculate the optical depth from the transmission. If we assume the dust particles to have a uniform mass and cross section, the column density can be calculated as
\begin{equation}
	\Sigma = \frac{\tau}{\sigma} m_p\,,
\end{equation}
\noindent where $\Sigma$ is the column density, $\tau$ the optical depth, $\sigma$ the effective cross section and $m_p$ the mass of one particle. Here we assume that $\lambda \ll a$, where $\lambda$ is the observational wavelength and $a$ is the particle radius. In this case Mie theory says that the effective cross section is twice the geometrical cross section, i.e. $\sigma = 2 \pi a^2$. Applying this the column density can also be expressed as
\begin{equation}
	\Sigma = \frac{2}{3}{\tau}{a}{\rho}\,,
    \label{eq:columndensity}
\end{equation}
\noindent where $\rho$ is the mean solid particle density. This assumption is approximately valid in our case, since the smallest particles in the HD 107146 system have $a$=2.7$\micron$ and the central wavelength of STIS observations is $\sim$0.6$\micron$ (see Table~\ref{tab:observationalparameters}). 
We assume a dust population with particles larger than a radiation pressure blow-out size of $a\sim$2.7 microns for silicate grains on circular orbits \citep{Ricci2015} and therefore we can assume these smaller grains are not present. As the average wavelength is shorter than 0.7$\micron$, since the STIS central wavelength lies close to the V-band central wavelength and the STIS CCD is more sensitive for the bluer part of the EM spectrum. 

It is important to notice that in order to calculate the amount of dust we still need to make assumptions on the particle size and density, however not on the albedo of the dust particles. This is an advantage of our method, compared to measuring the scattering by dust. 

\begin{figure}
	\centering
    \includegraphics[width = \columnwidth]{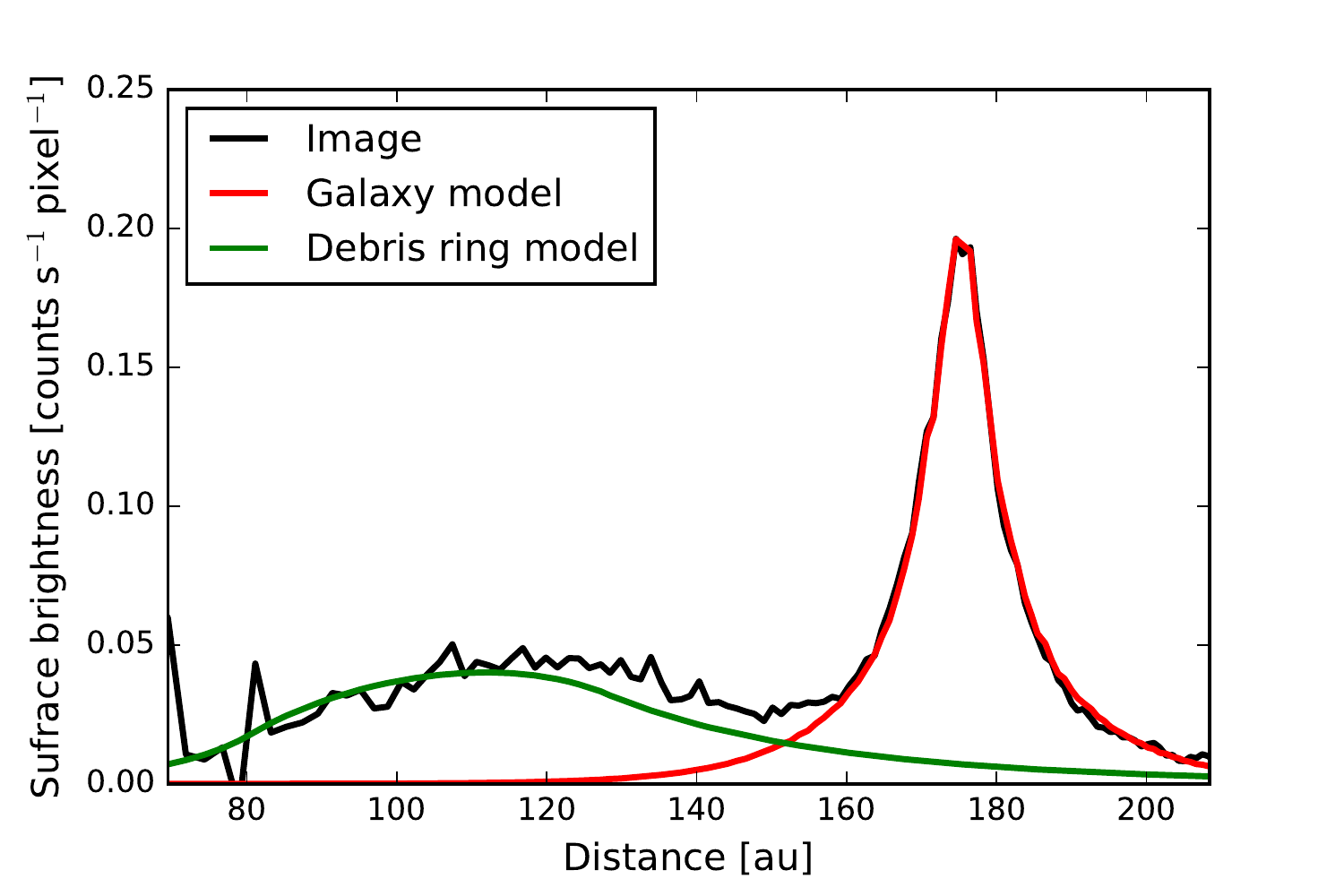}
	\caption{Average radial SB profile of the 2011 image of HD 107146 (see Figure~\protect\ref{fig:data2011}), the 2011 Vermin Galaxy model (see Section~\ref{s:model2011}) and the debris ring model (as described by \protect\cite{Schneider2016}) in the central elliptical sector (as indicated in Figure~\protect\ref{fig:overview}) as a function of the deprojected distance to the star. There clearly is a region (at a radius of $\sim$150 au) where both the debris ring and the galaxy contribute to the total surface brightness in the image.}
	\label{fig:surfacebrightness}
\end{figure}

\begin{figure*}
	\centering
    \includegraphics[width = 2\columnwidth]{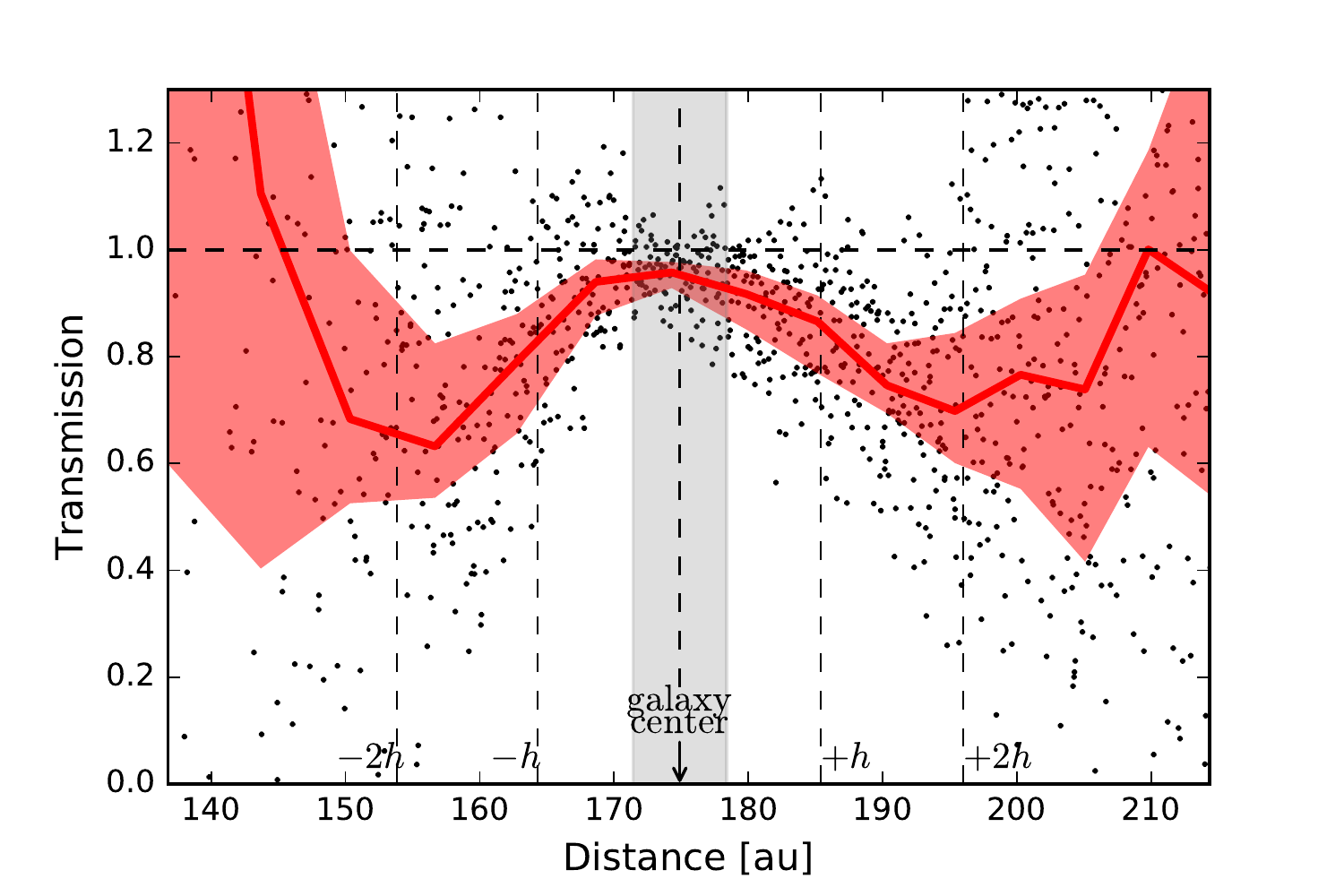}
    \caption{The radial transmission profile as a function of the deprojected distance in au to the star. The black data-points are the transmission value for each pixel within the central elliptical sector indicated in Figure~\ref{fig:overview}. The red line indicates 50$\%$-confidence level and the red shaded area indicates the (0.32, 0.68) confidence interval. The position of the galaxy center and the regions between one and two times the effective radius $h$ are indicated. The average transmission $\mu_{T'}$ and the corresponding standard deviation $\sigma_{T'}$ used to calculate the column density is determined in the gray region between $\sim$171 au and $\sim$178 au.}
    \label{fig:transmission}
\end{figure*}

\section{Results}
\label{s:results}

In order to apply the debris ring transit photometry method (as described in the previous section) effectively to the case of HD 107146 between the 2004/2001 epochs, it is naturally required that there is a region of spatial overlap between the SB-profiles of the debris ring and the background galaxy. 
In Figure~\ref{fig:data2011}, it seems that in the 2011 image the periphery of the HD 107146 debris disk is at the brink of transiting the background Vermin Galaxy; there may already be debris dust in between the line-of-sight of the observer and (parts of) the Vermin Galaxy. As can be seen in Figure~\ref{fig:surfacebrightness}, there is a region where both the background galaxy and the debris ring contribute flux to the total observed radial SB profile. We apply debris ring transit photometry to the 2011 image of HD 107146 in this overlapping region. 

{So far we have constructed our Vermin galaxy models with the assumption that in effect there is no attenuation at either epochs (2004 and 2011): the normalization of the galaxy mode $I_0$ was left as a free parameter. And because the instruments used are completely different in each epoch, one cannot draw conclusions regarding any differences in the fluxes observed in 2004 versus those of 2011. So our assumption has been there is no uniform screen of dust in front of the Vermin galaxy in 2011. 
We also attributed deviations from the galaxy model to intrinsic structure of the galaxy itself, not small attenuating structures, although that remains a possibility. 

The principal assumption in the overlapping galaxy method is that the far side (away from the overlap region) of the background galaxy is essentially unaffected by attenuation by the foreground object \citep{Holwerda2007,Holwerda2009,Holwerda2013}. If this assumption holds here i.e. that the part of the Vermin galaxy furthest away from HD 107146 on the sky, including the central parts, does not have significant attenuation in front of it, then the {\sc imfit} model is mostly based on unattenuated flux and should be representative of the flux missing closer to the disk.
To test this, we average the attenuation in bins defined by the projected radius from HD 107146. }

\subsection{The radial transmission profile}
\label{s:transmissionresults}

The transmission $T'$ of the Vermin Galaxy light through the debris disk can be obtained using Equation~\ref{eq:transmission}. Here the term $<D+Ge^{-\tau}>$ corresponds to the measured SB in the 2011 image (see Figure~\ref{fig:data2011}). $G'$ is the galaxy model as described in Section \ref{s:IMFIT} and $D'$ is the debris ring model by \cite{Schneider2016}. Figure~\ref{fig:transmission} shows the resulting radial transmission profile for the elliptical sector (as indicated in Figure~\ref{fig:overview}) where the debris disk and the galaxy overlap, which is the region around the center of the galaxy between $\sim$133-217 au projected distance from HD 107146. The transmission for each pixel within the elliptical sector is calculated and plotted in Figure \ref{fig:transmission}. The red shaded area indicates the (0.32, 0.68)-confidence region and the red line indicates the 50$\%$ confidence level for the transmission at a given radii.

These confidence levels are determined by dividing the region into 10 radial bins. For every bin we calculate the Empirical Distribution Function from which the estimates of the 32$\%$, 50$\%$ and 68$\%$ confidence levels are determined. This means the red region corresponds to the 1$\sigma$-uncertainty in the case when the distribution of the transmission follows a Gaussian distribution. The transmission profile in Figure \ref{fig:transmission} is still consistent with no attenuation within $3\sigma$ in any of the bins.

We note that the transmission seems to be $T'<$1, even in the central region of the galaxy. This is what one would naively expect to see when a debris disk attenuates light across the width of the galaxy, including the central and far side. However, this deviation from T=0 can also be caused by a bias in the background galaxy model used (see Section~\ref{s:discussion}). We note the lack of symmetry and conclude that the profile across the galaxy is suggestive of a remaining systematic error or offset in the galaxy model and not a steep dust gradient across it. 

\subsection{Column density}
\label{s:resultscolumndensity}

We aimed to determine the column density of dust by focusing on the dark gray region in Figure~\ref{fig:transmission} where the S/N is highest. The average transmission is determined in the region between $\sim$171 au and $\sim$178 au. We focus on the central part of the Vermin galaxy because this offers the highest galaxy flux and hence signal-to-noise. However, to counter this, there is some residual in the STIS fit that could well be remaining structure in the Vermin galaxy (see \S \ref{s:discussion}). 

With assumptions on particle size and density distributions and using Equation~\ref{eq:columndensity} the calculated average transmission can be converted into an average column density. We assume the grain size distribution and mean solid particle density which \cite{Ricci2015} used in their analysis of the HD 107146 debris disk based on mm wavelength imaging with ALMA.
They take a particle distribution $n$($a$)$ \sim a^{-q}$ where $a$ is the radius of the particle and $q$ the power law index. This distribution is valid between a specified minimal and maximal grain size ($a_{\rm{min}}$ and $a_{\rm{max}}$ respectively). In their paper it was assumed that $a_{\rm{min}} $=$ 2.7 \ \micron$ and $a_{\rm{max}} $=$ 2$ cm. They determine the value of $q$ from the spectral index of the dust opacity and the spectral index of small spherical interstellar grains using a relation derived by \cite{Draine2006} and found $q $=$ 3.25 \pm 0.09$. 

\begin{figure*}
	\centering
    \includegraphics[width = 2\columnwidth]{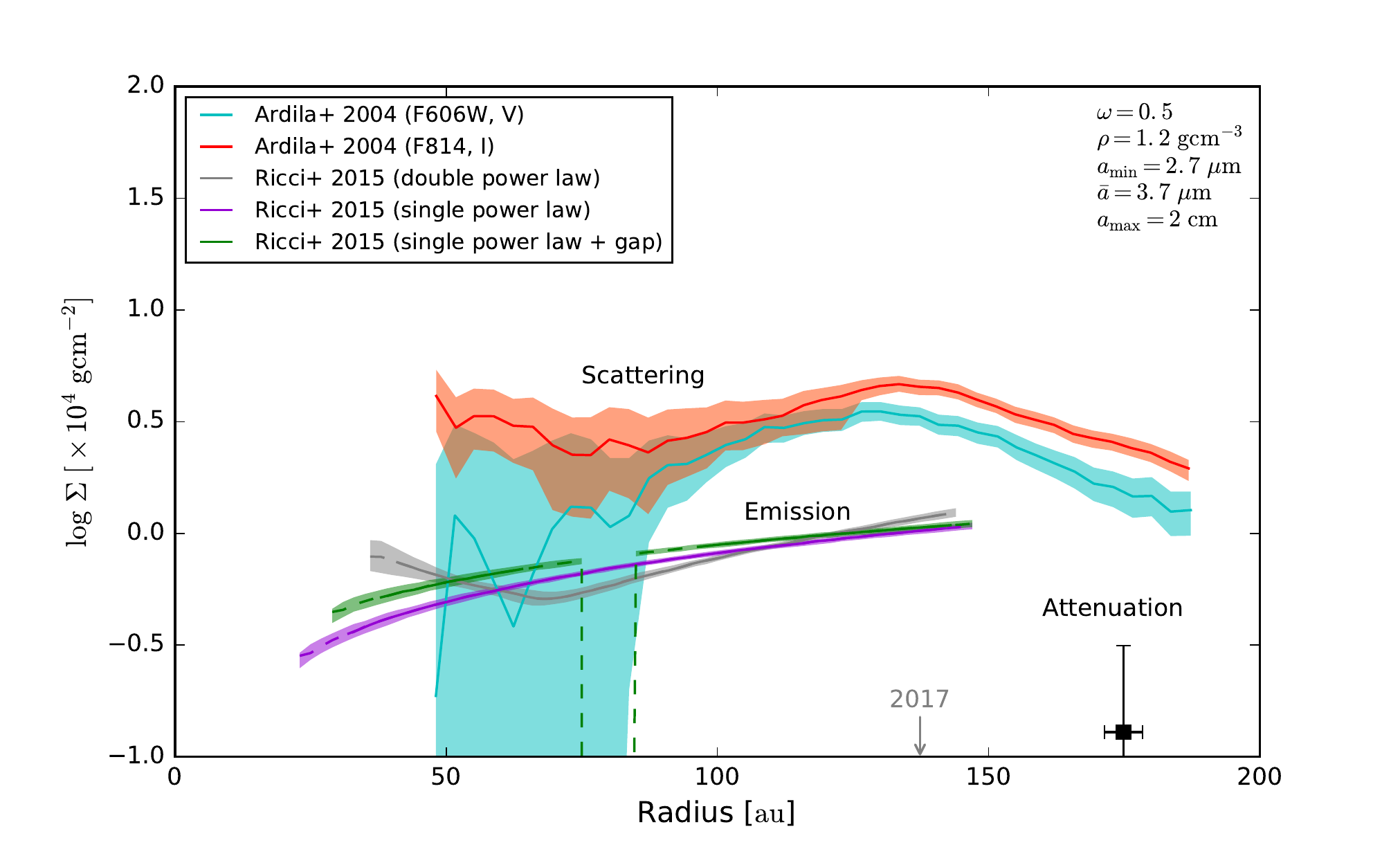}
    \caption{The line-of-sight column density profile of the HD 107146 debris disk as determined by ~\protect\cite{Ardila2004} (using our assumptions), \protect\cite{Ricci2015}, and the attenuation upper limit measure from the Vermin galaxy. These results use three different physical processes: scatter, emission, and attenuation, respectively. We made assumptions on the albedo $\omega$, mean solid particle density $\rho$, minimal particle radius $a_{\rm{min}}$, average particle radius $\bar{a}$ and maximal particle radius $a_{\rm{max}}$, which are shown in the right top corner (the assumptions are described in more detail in~Section~\ref{s:resultscolumndensity}). \protect\cite{Ardila2004} did observations in the V- and I-band. \protect\cite{Ricci2015} used three different models to explain the morphology of the debris disk: (1) a double power law (2) a single power law (3) a single power law with a gap. Our measurement of the average column density is taken in the region between $\sim$171-178 au from the star. The gray arrow indicates the position of the center of the galaxy in 2017 exactly six years after the 2011 observation, the first epoch of our follow-up campaign. All three detection methods have different sensitivities to particle size distribution and future work will hopefully constrain the particle distribution using all three measures.}
    \label{fig:columndensity}
\end{figure*}

We use the average particle size $\bar{a}$ of the distribution described above. This means the particle size for which the relation
\begin{equation}
	\int_{a_{\rm{min}}}^{\bar{a}} \! n({a'}) \, \mathrm{d}a' = \frac{1}{2}\,,
\end{equation}
\noindent holds, where $n(a)$ is normalized. Solving this relation gives $\bar{a} $=$ 3.7 \ \mu$m. We also assume the same mean solid particle density ($\rho$) as \cite{Ricci2015} of $\rho = 1.2 \ \rm{g} \ \rm{cm^{-3}}$.\\
Assuming $\bar{a} $=$ 3.7 \ \mu$m and $\rho $=$ 1.2 \ \rm{g cm^{-3}}$ we calculate the column density from the average transmission using Equation~\ref{eq:columndensity}. The result is plotted in Figure~\ref{fig:columndensity}.

In order to compare our column density with previous results on the debris disk of HD 107146, we also plot the column density determined by \cite{Ricci2015} in Figure~\ref{fig:columndensity}. Furthermore, the results of \cite{Ardila2004, Ardila2005} are included, where we convert their value $\tau \omega$ (optical depth $\times$ albedo) into a column density using Equation~\ref{eq:columndensity}. This was done by using the assumptions on the particle radius and mean solid particle density mentioned above and by assuming an albedo of $\omega = 0.5$ (an assumption which was also used by \cite{Ricci2015}). We note however that the three different methods are not sensitive to the same particle sizes and thus are likely to disagree once converted to a column density. 

\section{Comparison with previous research}
\label{s:comparison}

The debris disk around HD 107146 has been studied extensively in both scattered light and emission \citep{Ardila2004,Ertel2011,Schneider2014,Ricci2015}. The Vermin galaxy was previously viewed as an complication for disk photometry but in this paper, we showed a possible application for this object as a known light source for attenuation measures.

Given the position of the Vermin Galaxy behind the HD 107146 debris disk ($\sim$175 au in scaled projection from the star in 2011,  Figure~\ref{fig:surfacebrightness}), any dust attenuating it can be expected in the outermost edge of the debris disk. We examine the region from $\sim$171 au to $\sim$178 au from the star (see Figure~\ref{fig:columndensity}). This is at a larger radius than the region where the mm-sized grain distribution is evident $\sim$150 au \citep{Ricci2015}, but where $\micron$-sized particles are still present \citep{Ardila2004, Ertel2011, Schneider2014}. 


In order to compare the different results on the debris disk around HD 107146, we convert all previous and our results to a column density. Figure~\ref{fig:columndensity} shows the column density profiles inferred from thermal emission of dust at millimeter wavelengths \citep{Ricci2015} or optical scattered light \citep{Ardila2004}, and the upper limit inferred from the V-band attenuation (as described in Section~\ref{s:radialprofile}). The order-of-magnitude discrepancies between the different methods points to the different sensitivities to different particle sizes (e.g., mm- and $\micron$-sized dust particles). 


To convert our attenuation into a column density, we assumed an average particle size. While there are a range of sizes present in the disk, the average is not expected to change much if the particle distribution indeed follows a power law $n$($a$)$ \sim a^{-q}$, where the smallest particles still contribute the most attenuation, an assumption that holds for steep size distributions. 


The implied transmission value nearer the circumstellar disk ($\sim$150 au) is lower, implying a higher dust surface density (Figure \ref{fig:transmission}) but the pattern of lower transmission repeats itself on the other side of the Vermin galaxy (at 180-200 au) strongly implicating our galaxy model as the source of the underestimate of the flux, not dust attenuation. We cannot rule out that the offset in transmission at 175 au is due to the difference in instruments between epochs, introducing a systematic error and one should treat the value in Figure \ref{fig:columndensity} as an upper limit. The new STIS observations can best use this existing STIS image as a point of reference.

\section{Discussion}
\label{s:discussion}

An important assumption we make is that the galaxy model based on the 2004 image is accurate (i.e. $G$ = $G'$). However, this implies that after the fitting procedure (see Section~\ref{s:IMFIT}) a perfect model of the galaxy is obtained. Figure~\ref{fig:imfit} shows that in the residual image there is still  some leftover structure. This indicates that the galaxy model ($G'$) gives a good but not perfect estimate for the real galaxy SB profile ($G$). The presence of the leftover structures could be due to star formation regions, or could be caused by a wrong estimation of one of the model parameters (possibly due to going from one camera to another). 

Detailed analysis of the 2011 residuals (see Figure~\ref{fig:imfit}) reveals the presence of a ring around the center of the galaxy which has a slightly lower SB. This structure is the cause of the symmetric shape in the radial transmission profile (see Figure~\ref{fig:transmission}), as described in Section~\ref{s:transmissionresults}. This could for example be due to an overestimation of the scale length $h$, since this would imply that the extension of the galaxy model is larger than for the real galaxy, therefore overestimating the brightness in the outer regions. Such a error in parameter estimation could potentially be caused by the influence of light of the debris ring upon the modeling process of the galaxy, since this structure is not clearly present in our 2004 residual (again see Figure~\ref{fig:imfit}). Rings of star-forming structures in disk galaxies are not uncommon but it would strain our capacity with {\sc imfit} to model this ring in the present data, given the small differences between the two epochs in instrument sampling, resolution and wavelength range.

\begin{figure}
	\centering
    \includegraphics[width = 1.04\columnwidth]{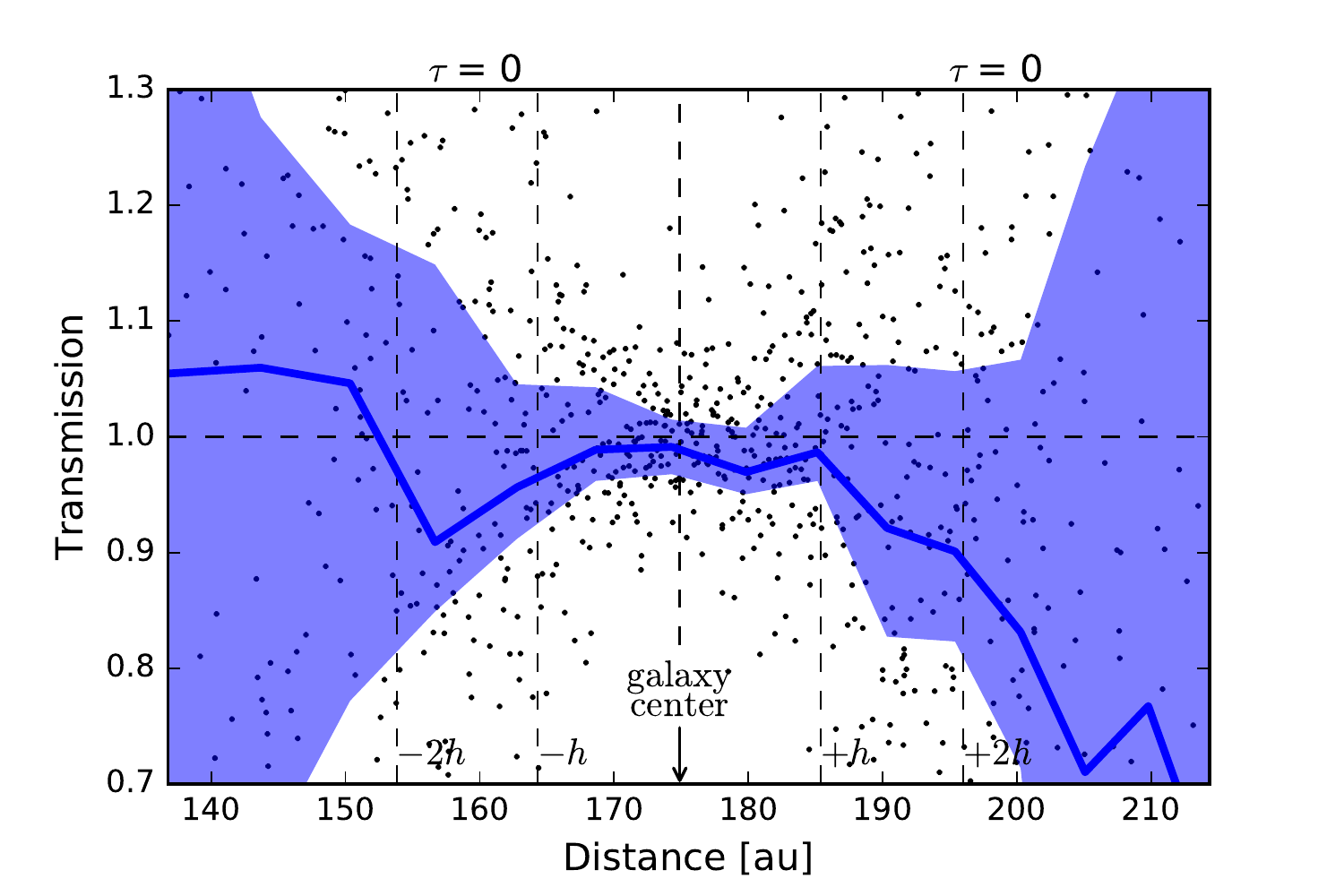}
    \includegraphics[width = 0.935\columnwidth]{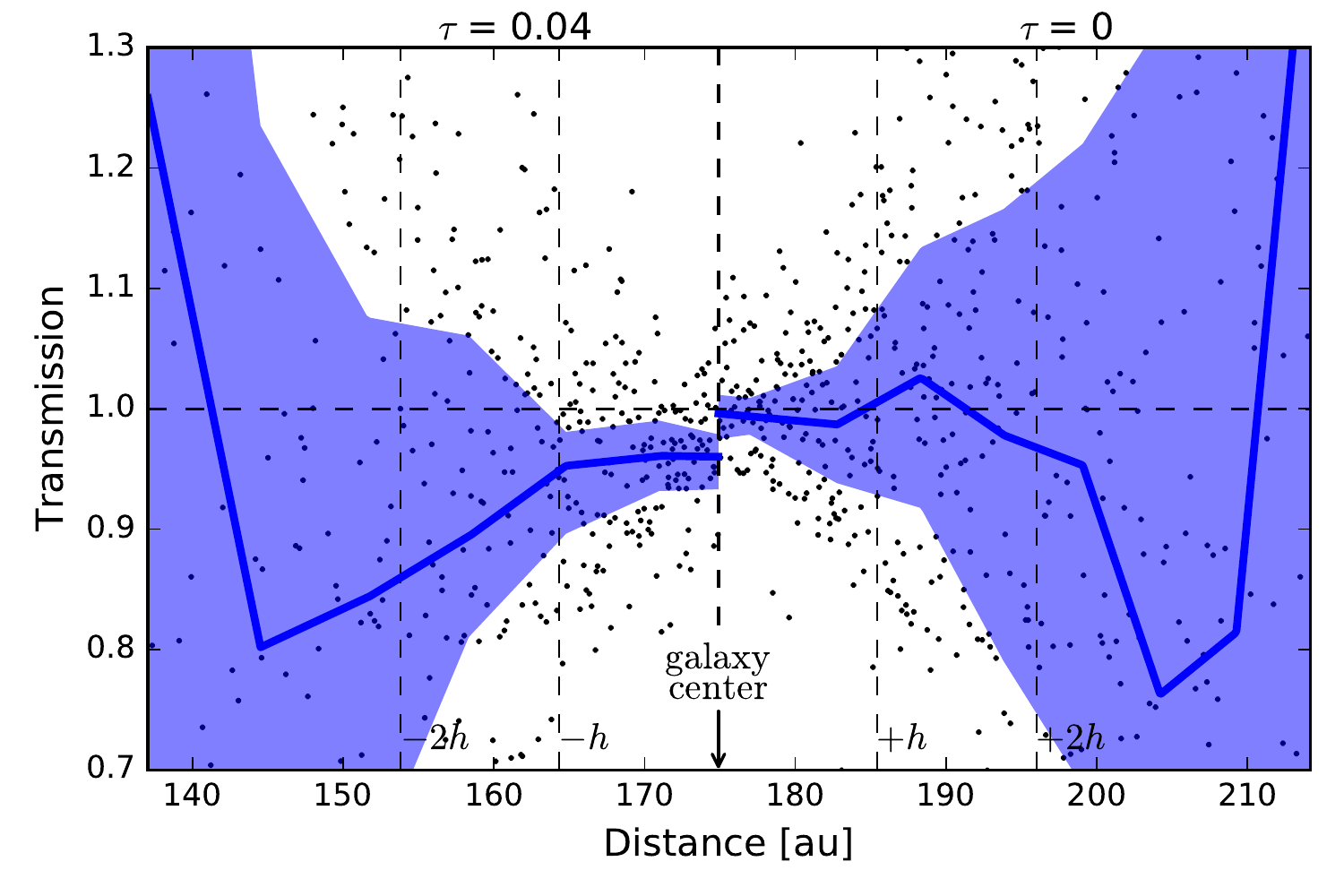}
    \caption{The transmission profile of an artificial image created by combining the galaxy model and the debris ring model and adding shot noise $\sigma$=0.005 to these models. \textit{Top image:} $\tau$ = 0 everywhere, as would be the case if no debris ring is present. \textit{Bottom image:} an uniform debris ring is present with $\tau$ = 0.04 which extends to the center of the galaxy and for larger radii $\tau$ = 0. The difference $\Delta \tau \geq 0.04$ can be detected for this noise value.}
    \label{fig:artificial}
\end{figure}

To check if the shape is indeed caused by a mismatch between the model and the data we tried the following approach. From our galaxy model $G'$ and our debris ring model $D'$, an artificial image $G'\rm{e}^{-\tau(r)} + D'$ is created where $\tau(r)$ simulates the optical depth of a debris disk that occults the light from the galaxy. Gaussian distributed noise $\sigma$ is added to both models, to simulate the effect of a finite exposure time with photon shot and readout noise. Here we set $\sigma$ = 0.005 counts/second/pixel which matches the noise in a background region of the HST images at a comparable stellocentric angle. We examine the transmission profile which is generated for the case without the presence of a debris ring (i.e.$\tau(r)$ = 0 everywhere), uncertainties estimated as in Figure \ref{fig:transmission}. The result is shown in the top image of Figure~\ref{fig:artificial}. In the central regions of the galaxy the transmission is centered around $T$ = 1, which we expect when there is no dust in front of the galaxy. Furthermore, this generated transmission profile shows that the error bars grow rapidly for larger galaxy radii where the flux from the galaxy drops below the noise level in the image. It gives us a useful range of points in the Vermin galaxy (within a scale-length) for reliable transmission measurements in future STIS observations.

The transmission profile with symmetry around the galaxy's center is not present in these simulations, suggesting that this profile indeed is caused by a systematic effect present in the outer regions of the galaxy model ($r>h$) not accurately representing the surface brightness profile.


The 2011 epoch STIS Vermin galaxy image, however can be used to determine the dust distribution in the debris ring if combined with planned multi-epoch HST/STIS observations of HD 107146 with an identical instrument setup (see Section~\ref{s:future}). These images can be used to directly measure the ratio of the fluxes corresponding to the same locations in the galaxy i.e. without going through the intervening step of generating a model image. If the initial (2011) flux measurement is defined as $F_1$ and the new measurement $F_2$, this ratio will be equal to
\begin{equation}
	{\frac{F_2}{F_1} = \frac{Ge^{-\tau_2}+D_2}{Ge^{-\tau_1}+D_1}}\,,
\end{equation} \noindent
where $G$ is the flux from the galaxy and the flux of the debris ring is given by $D_1$ and $D_2$ respectively. When the models of the debris disk $D_1$ and $D_2$ are removed from the measurements $F_1$ and $F_2$ first, this gives a direct measurement of the ratio of the transmissions $T_2/T_1 = e^{-\Delta \tau}$ where $\Delta \tau = \tau_2 - \tau_1$ (see Equation~\ref{eq:deftransmission}). 

The disadvantage is that this approach measures the attenuation difference with the 2011 epoch ($\Delta \tau$) rather than $\tau$ itself or needs to assume no attenuation in 2011.
The Vermin galaxy has not been occulted by the bright parts of the debris ring and therefore $\Delta \tau \approx \tau_2$ is likely to be a good approximation (Figure~\ref{fig:data2011}).

In order to estimate how much of a differential transmission one could detect with two STIS observation epochs, we generated an artificial image from the debris ring model and the galaxy model, with the same noise $\sigma = 0.005$, but now with $\tau(r)$>0 to mimic future STIS observations (\S \ref{s:future}).
Trial and error shows that for $\Delta \tau = 0.04$ there is a $1 \sigma$ detection in the transmission of the inner ($r<h$) region of the Vermin galaxy. The generated transmission profile for this value is shown in the bottom image of Figure~\ref{fig:artificial}. We place a $\tau = 0.04$ debris disk to extend out from the star halfway occulting the Vermin galaxy to better illustrate the effect. With the $1\sigma$ limit to attenuation change of $\Delta \tau = 0.04$ , averaged over one half of the Vermin galaxy, at least one reliable measurement of the attenuation should be possible for each of the future epochs by averaging over the whole or a large fraction of the Vermin galaxy.


In our view, three improvements over the current analysis can be made with the now in-progress STIS campaign on the ring-transit of the Vermin galaxy. The first improvement is that the 2011 image of the Vermin galaxy is used directly as a reference for transit photometry \citep[similar to rotated images in the occulting galaxy technique][]{Holwerda2007}. 
And secondly, an improved model of the Vermin galaxy using multiple epochs of STIS data may be made. If there is indeed a ring structure in the Vermin galaxy, a new {\sc imfit} model\footnote{{\sc imfit} is continuously improved and has added features that may allow an improved model and uncertainties in the near future. A development version could for example fit the STIS and ACS data --both filters--simultaneously.} could incorporate one to generate a noiseless image as a photometric reference.
Thirdly, an ensemble of attenuation values, taken at different epochs, can be compared to sub-mm emission and optical scatter observations to constrain the grain size distribution and average optical properties such as albedo.

\section{Conclusions}
\label{s:conclusions}
We conclude the following from this study:

\begin{itemize}
\item[I]{From out-of-transit observations from 2004 by HST/ACS, the Vermin Galaxy is a distant, smooth spiral galaxy that can be modeled with an exponential disk and a S\'{e}rsic pseudo-bulge in the V- and I-band.}
\item[II]{Our best Vermin Galaxy model based on the edge-on-transit HST/STIS observations from 2011 is insufficient to perform debris ring transit photometry with just these two epochs.}
\item[III]{Future transit observations by HST/STIS combined with the 2011 observations will be able to detect fluctuations at $1\sigma$ in the dust distribution of $\Delta \tau \geq 0.04$, providing an independent constraint on dust opacity in a debris disk.}
\end{itemize}

\section{Future research}
\label{s:future} 

Our results show that debris ring transit photometry is feasible using a galaxy as an extended background object, thus providing a measurement of the geometric opacity of dust in a foreground object. The brightest part of the debris disk around HD 107146 will move in front of a background galaxy (the Vermin galaxy) over the next decade. In the near future the Vermin Galaxy will be occulted by the inner parts of the debris disk, including where ALMA detects an inner ring of $\sim$mm size particles. These future positions are marked in Figure~\ref{fig:data2011} and~\ref{fig:columndensity}. 
Eventually in 2060, the transit of the galaxy behind the disk will be complete \citep{Zeegers2014}, but the Vermin Galaxy will cross the inner part of the debris disk first, where the residual halo of starlight will dominate any measurements. Therefore, near-future observations with Hubble Space Telescope will be our best chance to map the dust in the HD 107146 debris disk directly using transit photometry with the Vermin Galaxy. Observations at two different wavelengths could in principle constrain the grain size distribution, and to this end, new HST observations (HST Program 14714, P.I. G. Schneider) will be able to confirm dust densities if they are greater than $\Delta \tau=0.04$ with respect to the 2011 epoch.

Starting with the 2017 observations, the observing methodology is specifically optimized for imaging with higher signal-to-noise ratio in the region of the Vermin galaxy.  This was not the case in 2011 where the imaging program was designed to produce high fidelity SB image of the full extent and structure debris ring.

The new program will rely on a differential measure with respect to the 2011 baseline over six epochs, spaced by approximately six months, starting in 2017, as the galaxy transits behind the ring (Figure \ref{fig:data2011}).

\section*{Acknowledgements}

All the authors would like to emphatically thank the anonymous referee for their time, insight shared in meticulous and helpful comments that led to substantial improvements of the manuscript.

This study is based on observations made with the NASA/ ESA HST, obtained at the STScI, which is operated by the AURA, Inc., under NASA contract 5-26555. These observations are associated with programs 9987, 10330, and 12228. GS acknowledges support for programs 12228 and 14714 provided by NASA through a grant from STScI.

We would like to thank \cite{Ardila2004} and  \cite{Ricci2015} for their research on the dust particles in the debris disk around HD 107146, which was useful to compare with our own results. Furthermore, we want to thank \cite{Ardila2004} also for making their processed data available for further research. We would like to thank \cite{Erwin2015a} for developing the excellent astronomical image-fitting program  {\sc imfit}, and Dr. Erik Deul at Leiden Observatory for making it possible for us to use  {\sc imfit}. Moreover, we want to thank Dr. Michiel Hogerheijde at Leiden Observatory for helping us with the interpretation of our results.

\bibliographystyle{mnras} 
\bibliography{Bibliography}

\bsp	
\label{lastpage}
\end{document}